\documentclass[graybox]{svmult}

\usepackage{mathptmx}       
\usepackage{helvet}         
\usepackage{courier}        
\usepackage{type1cm}        
\usepackage{makeidx}         
\usepackage{graphicx}        
\usepackage{multicol}        
\usepackage[bottom]{footmisc}

\makeindex             

\usepackage{amsmath,amsfonts,amssymb,amscd,amsbsy,epsf}
\usepackage{cite}
\def\D{{\mathcal D}}
\def\F{{\mathcal F}}
\def\L{{\mathcal L}}
\def\real{{\mathbb R}}
\def\ep{\varepsilon}

\def\tp{\mathrm{p}}

\begin{document}
\title*{Generalized kinetic
Maxwell type models of granular gases}
\titlerunning{Generalized kinetic models
of granular gases}
\author{A.V. Bobylev, C. Cercignani and I.M. Gamba}
\institute{A.V. Bobylev \at Department of  Mathematics, Karlstad
University, Karlstad, SE-651\,88 Sweden.,
\email{alexander.bobylev@kau.se} \and C. Cercignani \at Politecnico
di Milano, Milano, Italy \email{carcer@mate.polimi.it}\and I.M.
Gamba \at Department of Mathematics, The University of Texas at
Austin, Austin, TX 78712-1082 U.S.A., \email{gamba@math.utexas.edu}
}
%
%
\maketitle

\begin{abstract} {  }
\end{abstract}


\section{Introduction}\label{sec:1}
It has been noticed in recent years that a significant non-trivial
physical phenomena in granular gases can be described mathematically
by dissipative Boltzmann type equations, as can be seen in
\cite{ba-tri-er} for a review in the area. As motivated by this
particular phenomena of energy dissipation at the kinetic level, we
consider in this chapter the Boltzmann equation for non-linear
interactions of Maxwell type and some generalizations of such
models.

The classical conservative (elastic) Boltzmann equation with the
Maxwell-type interactions is well-studied in the literature (see
\cite{B88,Ce} and references therein). Roughly speaking, this is a
mathematical model of a rarefied gas with binary collisions such
that the collision frequency is independent of the velocities of
colliding particles, and even though the intermolecular potentials
are not of those corresponding to hard sphere interactions, still
these models provide a very rich inside to the understanding of
kinetic evolution of gases.

Recently, Boltzmann equations of  Maxwell type were introduced for
models of granular gases were introduced in \cite{BCG} in three
dimensions, and a bit earlier in \cite{B-N-K} for in one dimension
case. Soon after that, these models became very popular among the
community studying granular gases (see, for example,
 the book \cite{PB} and
references therein). There are two obvious reasons for such studies
 The first one is that the
inelastic Maxwell-Boltzmann equation can be essentially simplified
by the Fourier transform similarly as done for the elastic case,
where its study becomes more transparent \cite{Bo76,BCG}. The second
reason is motivated by the special phenomena associated with {\em
homogeneous cooling} behavior, i.e.   solutions to the spatially
homogeneous inelastic Maxwell-Boltzmann equation have a non-trivial
self-similar asymptotics, and in addition,  the corresponding
self-similar solution has a power-like tail for large velocities.
The latter property was conjectured in \cite{E-B}
 and later proved in
\cite{BC-03,BCT-03}. This is a rather surprising fact, since the
Boltzmannn equation for hard spheres inelastic interactions has been
shown to have self similar solutions with all moments bounded and
large energy tails decaying exponentially.
 The conjecture of self-similar (or homogeneous cooling) states
 for such model of Maxwell type interactions was initially based
 on an exact $1 d$ solution constructed in \cite {BaMaPu}. It is
remarkable that such an asymptotics is absent in the elastic case
(as the elastic Boltzmann equation has {\em too many conservation
laws}). Later, the self-similar asymptotics was proved in the
elastic case for initial data with infinite energy \cite{BC-02} by
using another mathematical tools compared to \cite{BC-03} and
\cite{BG-06}.

 Surprisingly, the recently published exact self-similar solutions \cite{BG-06}
for elastic Maxwell type model for a slow down process, derived as a
formal asymptotic limit of a mixture,  also is shown to have
power-like tails. This fact  definitely suggests that self-similar
asymptotics are related to total energy dissipation rather than
local dissipative interactions. As an illustration to this fact, we
mention some recent publications \cite{BN05,To-05}, where $1 d$
Maxwell-type models were introduced for non-standard applications
such as  models in economics and social interactions, where also
self-similar asymptotics and power-like tail asymptotic states were
found.

Thus all the above discussed models describe qualitatively different
processes in physics or even in  economics, however their solutions
have a lot in common from mathematical point of view. It is also
clear that some further generalizations are possible: one can, for
example, include in the model multiple (not just binary)
interactions still assuming the constant (Maxwell-type) rate of
interactions. Will the multi-linear models have similar properties?
The answer is {\em yes}, as we shall see below.

Thus, it becomes clear that there must be some general mathematical
properties of Maxwell models, which, in turn, can explain properties
of  any particular model. That is to say there must be just one {\sl
main theorem}, from which one can deduce all above discussed facts
and their possible generalizations. Our goal is to consider Maxwell
models from very general point of view and to establish their key
properties that lead to the self-similar asymptotics.

All the results presented in this chapter  are mathematically
rigorous. Their full proofs can be found in \cite{BoCeGa}.

After this introduction, we introduce in Section~\ref{sec:2} three
specific Maxwell models of the Boltzmann equation: {\bf (A)}
classical (elastic) Boltzmann equation; {\bf (B)} the model (A) in
the presence of thermostat; {\bf (C) } inelastic Boltzmann equation
for Maxwell type interactions. Then, in Section~\ref{sec:3}, we
perform the Fourier transform and introduce an equation that
includes all the three models as particular cases. A further
generalization is done in Section~\ref{sec:4}, where the concept of
generalized multi-linear Maxwell model (in the Fourier space) is
introduced. Such models and their generalizations are studied in
detail in Sections~\ref{sec:5} and ~\ref{sec:6}. The most important
for our approach concept of L-Lipschitz nonlinear operator is
explained in Section~\ref{sec:4}. It is shown (Theorem 4.2) that all
multi-linear Maxwell models satisfy the L-Lipschitz condition. This
property of the models constitutes a basis for the general theory.

 The existence and uniqueness of solutions to the initial value
problem is stated in Subsection~\ref{sec:5.1} (Theorem 5.2). Then,
in Subsection~\ref{sec:5.2}, we present and study
 the large time asymptotics under very general conditions that are fulfilled, in
particular, for all our models. It is shown that {\em L-Lipschitz}
condition leads to self-similar asymptotics, provided the
corresponding self-similar solution does exist. The existence and
uniqueness of such self-similar solutions is stated in
Subsection~\ref{sec:5.3} (Theorem 5.12). This theorem  can be
considered, to some extent, as the {\sl main theorem}  for general
Maxwell-type models. Then, in Subsection~\ref{sec:5.4},
 we go back to
multi-linear models of Section~\ref{sec:4} and study more specific
properties of their self-similar solutions.

We explain in Section~\ref{sec:6} how to use our theory for
applications to any specific model: it is shown that the results can
be expressed in terms of just one function $\mu (p), p>0,$ that
depends on spectral properties of the specific model. General
properties (positivity, power-like tails, etc.) self-similar
solutions are studied in Subsections~\ref{sec:6.1} and
\ref{sec:6.2}. It includes also the case of $1 d$ models, where the
Laplace (instead of Fourier) transform is used. In Subsection 6.3,
we formulate,  in the unified statement (Theorem 11.1),  the main
properties of Maxwell models {\bf (A),(B)} and {\bf (C)} of the
Boltzmann equation. This result is, in particular, an essential
improvement of earlier results of [7] for the model {\bf (A) } and
quite new for the model {\bf (B).}

Applications to one dimensional models are also briefly discussed at the end
of Subsection 6.3.

\section{Maxwell models of the Boltzmann equation}\label{sec:2}
 We consider a spatially homogeneous rarefied $d$-dimensional
 gas $(d = 2,3,\ldots)$ of particles having a unit mass.
 Let $f(v,t)$, where $v\in\real^d$ and $t\in\real_+$ denote respectively
 the velocity and time variables, be a  one-particle distribution
 function with usual normalization
 \begin{equation}\label{eq2.1}
 \int_{\real^d} dv\, f(v,t) = 1\ .
 \end{equation}
 Then $f(v,t)$ has an obvious meaning of a time-dependent
 probability density in $\real^d$.
 We assume that the collision frequency is independent of the velocities
 of the colliding particles (Maxwell-type interactions). We  discuss
three  different physical models (A), (B) and (C).

 {\bf (A)  Classical Maxwell gas (elastic collisions).}
 In this case  $f(v,t)$ satisfies the usual Boltzmann equation
 \begin{equation}\label{eq2.2}
f_t = Q(f,f)
 = \int_{\real^d \times S^{d-1}}
 \mkern-32mu   dw\,d\omega\, g
 ( \frac{u\cdot \omega}{|u|} )
 [f(v')f(w') - f(v)f(w)]\ ,
 \end{equation}
 where the exchange of velocities after a collision are given by
$$ v' = \frac12 (v+w+|u|\omega)\, ,\qquad \ \text{and}\ \ \quad
  w' = \frac12 (v+w - |u|\omega)
$$ where
 $u= v-w$ is the relative velocity and $\Omega \in S^{d-1}$.
 For the sake of brevity we shall consider below the model
 non-negative collision kernels $g(s)$ such that $g(s)$ is integrable
 on $[-1,1]$.
 The argument $t$  of $f(v,t)$ and similar functions is often omitted
 below (as in Eq.~\eqref{eq2.2}).

 {\bf (B) Elastic model with a thermostat}
 This case corresponds to  model  {\bf (A) }
in the presence of a thermostat  that consists of
 Maxwell particles with mass $m>0$  having the Maxwellian
 distribution
 \begin{equation}\label{eq2.3}
 M(v) = ( \frac{2\pi T}{m})^{-d/2}
 \exp ( -\frac{m|v|^2}{2T})
 \end{equation}
 with a constant temperature $T>0$.
 Then the evolution equation  for $f(x,t)$ becomes
 \begin{equation}\label{eq2.4}
 f_t = Q(f,f) +\theta \int dw\,d\omega\, g
 (\frac{u\cdot \omega}{|u|})
[ f(v') M(w') - f(v) M(w)]\ ,
 \end{equation}
 where  $\theta >0$ is a coupling constant, and the exchange of velocities
is now
$$
v' = \frac{v+m(w+|u|\omega)}{1+m},\qquad \ \text{and}\ \ \quad
 w' = \frac{v+mw - |u|\omega}{1+m},
$$
with  $u= v-w$ the relative velocity and ,
 $\omega \in S^{d-1}$.

 Equation~\eqref{eq2.4} was derived in \cite{BG-06} as a
 certain limiting case of a binary mixture of weakly interacting
 Maxwell gases.

{\bf (C)  Maxwell model for inelastic particles.}
 We consider this model in the form given in \cite{BC-03}.
 Then the inelastic Boltzmann equation in the weak form reads
 \begin{equation}\label{eq2.5}
 \frac{\partial}{\partial t} (f,\psi)
 = \int_{\real^d\times \real^d \times S^{d-1}} \mkern-36mu
 dv\, dw\, d\omega\, f(v) f(w)
 \frac{|u\cdot\omega|} {|u|}
 [\psi (v')-\psi (v)]\ ,
 \end{equation}
 where $\psi (v)$ is a bounded and continuous test function,
 \begin{equation}\label{eq2.6}
 (f,\psi) = \int_{\real^d} \mkern-12mu dv\, f(v,t)\psi (v),\
 u= v-w,\ \omega \in S^{d-1}, \
 v' = v-\frac{1+e}{2} (u\cdot \omega) \omega\ ,
 \end{equation}
 the constant parameter $0 < e\le 1$ denotes the restitution coefficient.
 Note that the model {\bf (C}  with $e=1$ is equivalent to the model {\bf (A)} with some kernel $g(s)$.

 All three models can be simplified (in the mathematical sense) by
 taking the Fourier transform.

 We denote
 \begin{equation}\label{eq2.7}
 \hat f (k,t) = \F [f] = (f, e^{-ik\cdot v}),\qquad k\in\real^d\ ,
 \end{equation}
 and obtain (by using the same trick as in \cite{Bo76} for the model
 {\bf (A)})  for all three models the following equations:
 \begin{equation}\label{eq2.8}
 \text{\bf (A)}\quad\quad\quad\quad\quad\quad \quad
 \hat f_t = \widehat Q (\hat f,\hat f)
 = \int_{S^{d-1}} \mkern-20mu d\omega\, g(\frac{k\cdot\omega}{|k|})
 [\hat f(k_+) \hat f(k_-) - \hat f(k) \hat f(0)]\ ,\quad\quad\quad\quad
 \end{equation}
 where
 $k_\pm = \frac12 (k\pm |k|\omega)$, $ \omega \in S^{d-1}$,
$ \hat f(0)=1$.
 \begin{equation}\label{eq2.9}
 \text{\bf (B)}\quad \quad\quad\quad\quad\quad
 \hat f_t = \widehat Q (\hat f,\hat f) + \theta \int_{S^{d-1}}\mkern-20mu
 d\omega\, g(\frac{k\cdot\omega}{|k|})
 [\hat f(k_+) \widehat M (k_-) - \hat f(k) \widehat M(0)]\ ,\quad\quad\quad
 \end{equation}
 where
 $\widehat M (k) = e^{-\frac{T|k|^2}{2m}}$,
 $k_+ = \frac{k+m|k|\omega}{1+m}$,
 $k_- = k-k_+$,
$ \omega \in S^{d-1}$,
 $\hat f(0)=1$.
 \begin{equation}\label{eq2.10}
 \text{\bf (C)}\quad\quad\quad\quad\quad\quad \quad\quad\quad\quad\quad
 \hat f_t = \int_{S^{d-1}}\mkern-20mu
 d\omega \frac{|k\cdot\omega|}{|k|}
 [\hat f(k_+) \hat f(k_-) - \hat f (k) \hat f(0)]\ ,\quad\quad\quad\quad
\quad \quad\end{equation}
 where $\hat f (0) =1$,
 $k_+ = \frac{1+e}2 (k\cdot \omega)\omega$,
 $k_- = k-k_+$,   with
 $\omega \in S^{d-1}$ is the direction  containing the two centers of the particles at the
 time of the interaction. Equivalently, one may alternative write  $k_-=
 \frac{1+e}{4}(k-|k|n\tilde\omega),$ and
 $k_+ = k-k_-$, where now $\tilde\omega \in S^{d-1}$ is the direction of the post collisional relative
 velocity,  and  the term $\frac{|k\cdot\omega|}{|k|}\, dw$ is
 replaced by a function $g(\frac{k\cdot\tilde\omega}{|k|})d\tilde\omega.$

Case {\bf (B)} can be simplified by the  substitution
 \begin{equation}\label{eq2.11}
 \hat f(k,t) = \widetilde{\widehat  f} (k,t) \exp [ -
\frac{T|k|^2}2]\ \, ,
 \end{equation}
 leading, omitting tildes,  to the  equation
 \begin{equation}\label{eq2.12}
 \text{\bf (B$'$)}\quad \quad\quad\quad \quad\quad\quad
 \hat f_t = \widehat Q (\hat f,\hat f) + \theta \int_{S^{d-1}}\mkern-24mu
 d\omega\, g( \frac{k\cdot\omega}{|k|})
 [ \hat f( \frac{k+m|k|\omega}{1+m}) - \hat f(k)
 ]\ ,
\quad\quad \quad
 \end{equation}
 i.e., the model for {\bf (B)} with $T=0$, or equivalently a linear collisional term
 the background
 singular distribution.
 Therefore, we shall consider below just the case {\bf (B$'$)},
 assuming nevertheless
 that $\hat f(k,t)$ in Eq.~\eqref{eq2.12} is the Fourier transform
 \eqref{eq2.7}
 of a probability density $f(v,t)$.


\section{Isotropic Maxwell model in the Fourier representation}\label{sec:3}
We shall see that these three models {\bf  (A), (B)} and {\bf  (C)}
admit a class of isotropic solutions with distribution functions $f=
f(|v|,t)$. Indeed, according to   \eqref{eq2.7} we look for
solutions $\hat f= \hat f(|k|,t)$ to  the  corresponding isotropic
Fourier transformed problem, given by
\begin{equation}\label{eq3.1}
 x= |k|^2\ ,\quad
 \varphi (x,t) = \hat f(|k|,t) = \F[f(|v|,t)]\ ,
\end{equation}
 where  $\varphi (x,t)$ solves the following initial value problem
\begin{equation}\label{eq3.2}
\begin{split}
 \varphi_t & = \int_0^1 ds\, G(s) \left\{ \varphi [a(s)x] \varphi [b(s)x]
 - \varphi (x)\right\} +\\
 &\qquad +
 \int_0^1 ds H(s) \left\{ \varphi [c(s)x] - \varphi (x)\right\}\ ,\\
&\ \text{}
\quad \varphi_{{t=0}} =\varphi_{{0}}(x)\, ,\  \ \  \text{}\ \  \varphi (0,t)=1\ ,
 \end{split}
\end{equation}
where $a(s)$, $b(s)$, $c(s)$ are non-negative continuous functions
on $[0,1]$, whereas $G(s)$ and $H(s)$ are generalized non-negative
 functions such that
\begin{equation}\label{eq3.3}
 \int_0^1 ds\, G(s) <\infty\ ,\qquad
 \int_0^1 ds\, H(s) < \infty\ .
\end{equation}
Thus, we do not exclude such functions as $G=\delta (s-s_0)$, $0<
s_0 <1$, etc. We shall see below that, for isotropic solutions
\eqref{eq3.1}, each of the three equations \eqref{eq2.8},
\eqref{eq2.10}, \eqref{eq2.12}
 is a particular case of Eq.~\eqref{eq3.2}.

Let us first consider Eq.~\eqref{eq2.8} with $\hat f (k,t) = \varphi
(x,t)$
 in the notation \eqref{eq3.1}.
 In that case
 $$|k_\pm|^2 = |k|^2 \frac{1\pm (\omega_0 \cdot\omega)}{2}\ ,\qquad
 \omega_0 = \frac{k}{|k|} \in S^{d-1}\ ,\  d=2,\ldots,$$
 and the integral in Eq.~\eqref{eq2.8} reads
 \begin{equation}\label{eq3.4}
 \int_{S^{d-1}} d\omega\, g (\omega_0\cdot \omega) \varphi
 \left[ x\frac{1+\omega_0 \cdot \omega}{2} \right] \varphi
 \left[ x\frac{1-\omega_0\cdot\omega}{2}\right]\  .
 \end{equation}
 It is easy to verify the identity
 \begin{equation}\label{eq3.5}
 \int_{S^{d-1}} d\omega\, F (\omega\cdot\omega_0)
 = |S^{d-2}| \int_{-1}^1 dz\, F(z) (1-z^2)^{\frac{d-3}2}\ ,
 \end{equation}
 where $|S^{d-2}|$ denotes the ``area'' of the unit sphere in $\real^{d-1}$
 for $d\ge 3$ and $|S^0 | =2$.
 The identity \eqref{eq3.5} holds for any function $F(z)$ provided the
 integral as defined  in the right hand side of \eqref{eq3.5} exists.

 The integral \eqref{eq3.4} now reads
 \begin{equation*}
 \begin{split}
 & |S^{d-2}| \int_{-1}^1 dz\, g (z) (1-z^2)^{\frac{d-3}2}
 \varphi ( x\frac{1+z}2) \varphi ( x\frac{1-z}2) =\\
 &\qquad = \int_0^1 ds\, G(s)\varphi (sx) \varphi [(1-s)x]\ ,
 \end{split}
 \end{equation*}
 where
 \begin{equation}\label{eq3.6}
 G(s) = 2^{d-2} |S^{d-2}| g(1-2s) [s(1-s)]^{\frac{d-3}2}\ ,\qquad
 d= 2,3,\ldots\ .
 \end{equation}

 Hence, in this case we obtain Eq.~\eqref{eq3.2}, where
 \begin{equation}\label{eq3.7}
 \text{\bf (A)}\qquad \qquad \qquad  \qquad
 a(s) =s\ ,\qquad b(s) = 1-s\ ,\qquad H(s)=0\ ,\quad\quad \quad
\qquad \quad\quad \quad
 \end{equation}
 $G(s)$ is given in Eq.~\eqref{eq3.6}.

 Two other models  {\bf (B$'$)} and {\bf (C)}, described by
eqs.~\eqref{eq2.12},
 \eqref{eq2.10} respectively, can be considered quite similarly.
 In both cases we obtain Eq.~\eqref{eq3.2}, where
 \begin{equation}\label{eq3.8}
 \begin{split}
 \text{\bf (B$'$)}\qquad \qquad \qquad  \qquad  \qquad
 &a(s) =s,\quad
 b(s) = 1-s,\quad
 c(s) = 1- \frac{4m}{(1+m)^2} s,\qquad \quad\quad\quad \\
 &H(s) = \theta G(s)\ ,
 \end{split}
 \end{equation}
 $G(s)$ is given in Eq.~\eqref{eq3.6}:
 \begin{equation}\label{eq3.9}
 \begin{split}
 \text{{\bf (C)}} \qquad \qquad \qquad \qquad
& a(s) = \frac{(1+e)^2}4 s,\quad
 b(s) =1 - \frac{(1+e)(3-e)}{4}\, s,\qquad\qquad \quad\quad\quad \\
 \noalign{\vskip6pt}
& H (s)=0,\quad
 G(s) = |S^{d-2}| (1-s)^{\frac{d-3}2}\ .
\end{split}
 \end{equation}

 Hence, all three models are described by Eq.~\eqref{eq3.2} where
 $0<a(s) ,b(s) ,c(s)\le 1$ are non-negative linear functions.
 One can also find in recent publications some other  useful equations that
 can be reduced after Fourier or Laplace transformations to
 Eq.~\eqref{eq3.2} (see, for example, \cite{BN05},
 \cite{To-05} that correspond to the case
 $G= \delta (s-s_0)$, $H=0$).

 The equation \eqref{eq3.2} with $H(s) =0$ first appeared in its general
 form in \cite{BC-03} in connection with models {\bf (A)}  and {\bf
(C)}.
 The consideration of the problem of self-similar asymptotics for
 Eq.~\eqref{eq3.2} in that paper made it quite clear that the most
 important properties of ``physical'' solutions depend very weakly on
the  specific functions $G(s)$, $a(s)$ and $b(s)$.

 \section{Models with multiple interactions}\label{sec:4}

We present now a general framework to study  solutions
to  the type of problems introduced in the previous section.

 We assume, without loss of generality, (scaling transformations
 $\tilde t = \alpha t$, $\alpha = \text{const.}$) that
 \begin{equation}\label{eq4.1}
 \int_0^1 ds\, [G(s) + H(s)] = 1
 \end{equation}
 in Eq.~\eqref{eq3.2}.
 Then Eq.~\eqref{eq3.2} can be considered as a particular case
 of the following equation for a function $u(x,t)$
 \begin{equation}\label{eq4.2}
 u_t + u = \Gamma (u)\ ,\qquad x\ge 0,\  t\ge 0\ ,
 \end{equation}
 where
 \begin{equation}\label{eq4.3}
 \begin{split}
 &\Gamma (u) = \sum_{n=1}^N \alpha_n \Gamma^{(n)} (u)\ ,\quad
 \sum_{n=1}^N \alpha_n =1\ ,\ \alpha_n \ge 0\ ,\\
& \Gamma^{(n)} (u) = \int_0^\infty\mkern-20mu  da_1  \ldots
\int_0^\infty \mkern-20mu da_n\
 A_n (a_1,\ldots, a_n) \prod_{k=1}^n  u(a_kx),\quad
 n=1,\ldots,N\ .
 \end{split}
 \end{equation}
 We assume that
 \begin{equation}\label{eq4.3bis}
A_n(a) = A_n (a_1,\ldots, a_n) \ge 0\ ,\qquad
 \int_0^\infty da_1 \ldots \int_0^\infty da_n\  A(a_1,\ldots,a_n) =1\ ,
 \end{equation}
 where  $A_n(a) = A_n (a_1,\ldots, a_n)$ is a
 generalized density of a probability
 measure in $\real_+^n$ for any $n=1,\ldots,N$.
 We also assume that all $A_n(a)$ have a compact support, i.e.,
 \begin{equation}\label{eq4.4}
 A_n(a_1,\ldots,a_n) \equiv 0\ \text{ if }\
 \sum_{k=1}^n a_k^2 > R^2\ ,\qquad
 n = 1,\ldots, N\ ,
 \end{equation}
 for sufficiently large $0< R<\infty$.

 Eq.~\eqref{eq3.2} is a particular case of Eq.~\eqref{eq4.2} with
 \begin{equation}\label{eq4.5}
 \begin{split}
 &N= 2\ ,\quad \alpha_1 = \int_0^1 ds\, H(s)\ ,\quad
 \alpha_2 = \int_0^1 ds\, G(s)\\
 &A_1 (a_1) = \frac1{\alpha_1} \int_0^1 ds\, H(s) \delta [a_1 - c(s)]\\
 &A_2 (a_1,a_2) = \frac1{\alpha_2} \int_0^1 ds\, G(s)
 \delta [a_1 -a(s)] \delta [a_2-b(s)]\ .
 \end{split}
 \end{equation}

 It is clear that Eq.~\eqref{eq4.2} can be considered as a
 generalized Fourier transformed isotropic Maxwell model with
 multiple interactions provided $u(0,t) =1$, the case $N=\infty$
 in Eqs.~\eqref{eq4.3} can be treated in the same way.

\subsection{Statement of
 the general problem}\label{sec:4.1}

 The general problem  we consider below can be formulated
 in the following way.
 We study the initial value problem
 \begin{equation}\label{eq4.6}
 u_t + u = \Gamma (u)\ ,\quad
 u_{|t=0} = u_0 (x)\ ,\quad
 x\ge 0\ ,\quad
 t\ge 0\ ,
 \end{equation}
 in the Banach space $B= C(\real_+)$ of continuous functions $u(x)$
 with the norm
 \begin{equation}\label{eq4.7}
 \| u\| = \sup_{x\ge 0} |u(x)|\ .
 \end{equation}

 It is usually assumed that $\|u_0\| \le 1$
and that the operator
 $\Gamma$ is given by Eqs. ~\eqref{eq4.3}.
 On the other hand, there are just a few properties of $\Gamma (u)$
 that are essential for existence, uniqueness and large time asymptotics
 of the solution $u(x,t)$ of the problem~\eqref{eq4.6}.
Therefore, in many cases  the results can be applied to more
general classes of operators $\Gamma$ in Eqs.~\eqref{eq4.6} and
more general functional space, for example $B= C(\real^d)$
(anisotropic models).
That is why we study below the class \eqref{eq4.3} of operators
$\Gamma$ as the most important example, but simultaneously indicate which
properties of $\Gamma$ are relevant in each case.
In particular, most of the  results of Section~4--6 do not use a specific form
\eqref{eq4.3} of $\Gamma$ and, in fact, are valid for a  more general class of
operators.

Following this way of study, we first consider the problem \eqref{eq4.6}
with $\Gamma$ given by Eqs.~\eqref{eq4.3} and point out the most
important properties of $\Gamma$.

We simplify notations and omit in most of the cases below the argument
$x$ of the function $u(x,t)$.
The notation $u(t)$ (instead of $u(x,t)$) means then the function of the  real
variable $t\ge 0$ with values in the space $B = C(\real_+)$.

\begin{remark}
We shall omit below the argument $x\in \real_+$ of functions $u(x)$, $v(x)$,
etc., in all cases when this does not cause a misunderstanding.
In particular, inequalities of the kind $|u| \le |v|$ should be
understood as a  point-wise  control in absolute value, i.e.
 ``$|u(x)| \le |v(x)|$ for any $x\ge 0$'' and so on.
\end{remark}

We first start by giving the following general definition for
 operators  acting on a unit ball  of a Banach space $B$ denoted by
\begin{equation}\label{eq4.10}
U = \{u \in B : \|u\| \le 1\}
\end{equation}

\medskip

\begin{definition}\label{def4.1} The operator  $\Gamma=\Gamma(u)$
is called an  $L$-Lipschitz operator if there exists a linear bounded
operator $L: B\to B$  such that
the inequality
\begin{equation}\label{eq5.5}
| \Gamma (u_1) - \Gamma (u_2)| \le L (|u_1 - u_2 |)
\end{equation}
holds for any pair of functions $ u_{1,2}$ in  $U $.
\end{definition}
\begin{remark}
Note that the $L$-Lipschitz condition  \eqref{eq5.5} holds, by definition,
at any point $x\in \real_+$ (or  $x\in \real^d$ if $B = C(\real^d)$).
Thus, condition
 \eqref{eq5.5} is much stronger than the classical Lipschitz condition
\begin{equation}\label{eq4.13}
\| \Gamma (u_1) - \Gamma (u_2) \|
< C \| u_1 - u_2\| \ \text{ if }\ \ u_{1,2} \in U
\end{equation}
which obviously follows from  \eqref{eq5.5} with the constant $C=\|L\|_B$,
the norm of the operator $L$ in the  space of bounded operators acting in $B$.
In other words, the terminology ``$L$-Lipschitz condition'' means the
point-wise Lipschitz condition with respect to an specific linear operator $L$.
\end{remark}
\medskip

 We assume, without loss of generality, that the kernels
$A_n (a_1,\ldots,a_n)$ in Eqs.~\eqref{eq4.3}
are symmetric with respect to any permutation of the arguments
$(a_1,\ldots,a_n)$, $n=2,3,\ldots,N$.

The next lemma states that the operator   $\Gamma(u)$ defined in
 Eqs.~\eqref{eq4.3}, which  satisfies $\Gamma(1)=1$ (mass conservation)
and  maps $U$ into itself,
 satisfies an   $L$-Lipschitz condition, where
 the
linear operator $L$ is the one given by the linearization of $\Gamma$
near  the unity. See \cite{BoCeGa} for its proof.

\begin{theorem}\label{theorem-lip}  The operator   $\Gamma(u)$ defined in
 Eqs.~\eqref{eq4.3} maps $U$ into itself  and
satisfies the $L$-Lipschitz condition \eqref{eq5.5},
where the linear operator  $L$ is given by
\begin{equation}\label{eq5.2}
Lu = \int_0^\infty da\, K (a) u (ax)\ ,
\end{equation}{with}
\begin{equation}\begin{split}\label{eq5.3}
K(a) = \sum_{n=1}^N n\alpha_n K_n (a)\, , \qquad\qquad\qquad\qquad\qquad\\
\text{where}\qquad K_n(a) = \int_0^\infty da_2\ldots \int_0^\infty
da_n \ A_n (a,a_2,\ldots,a_n) \ \ \text{and}\  \ \sum_{n=1}^N
\alpha_n =1\, .
\end{split}\end{equation}
for symmetric kernels $A_n(a,a_2,\ldots,a_n), n=2,\ldots N$\ .
\end{theorem}

And the following  corollary holds.

\begin{corollary}             
The Lipschitz condition \eqref{eq4.13} is fulfilled for $\Gamma (u)$
given in Eqs.~\eqref{eq4.3} with the constant
\begin{equation}\label{eq5.8}
C = \|L\| = \sum_{n=1}^N n\alpha_n\ ,\qquad
\sum_{n=1}^\infty \alpha_n =1\ ,
\end{equation}
where $\|L\|$ is the norm of $L$ in $B$.
\end{corollary}

 It can also be shown  that the $L$-Lipschitz condition holds in
 $B=C(\real^d)$ for ``gain-operators'' in Fourier transformed Boltzmann
equations  ~\eqref{eq2.8},  \eqref{eq2.9} and \eqref{eq2.10}.

\section{The general problem in Fourier  representation}\label{sec:5}

\vskip.5in
\subsection{Existence and uniqueness of solutions}\label{sec:5.1}
It is possible to show, with
 minimal requirements,  the existence and uniqueness
results associated with the initial value problem \eqref{eq4.6} in the Banach
 space $(B,\| \cdot \|)$,
 where the norm associated to B is defined in \eqref{eq4.7}
 In fact, this existence and uniqueness result is an
 application of the classical Picard iteration scheme
 and  holds for any  operator $\Gamma$
 which satisfies  the usual Lipschitz condition
\eqref{eq4.13} and  transforms the unit ball $U$ into itself. The
proof of all statements below can be found in \cite{BoCeGa}.

\begin{lemma}\label{lemma5.1} ({\sl Picard Iteration scheme})
 The operator   $\Gamma(u)$ maps $U$ into itself  and
satisfies the $L$-Lipschitz condition \eqref{eq4.13},
then the initial value
problem \eqref{eq4.6} with arbitrary $u_0 \in U$ has a
unique solution $u(t)$ such that $u(t) \in U$ for any $t\ge 0$.
\end{lemma}

Next, we observe that the $L$-Lipschitz condition yields an estimate
for  the difference of any two solutions of the problems in terms of
their initial states. This is a key fact in the development of the
further studies of the self similar asymptotics.

 \begin{theorem}\label{theorem5.2}
 Consider the Cauchy problem \eqref{eq4.6} with $\|u_0\| \le1$
 and assume that the operator $\Gamma :B\to B$
 \begin{itemize}
 \item[{\bf (a)}]
 maps the closed unit ball $U\subset B$ to itself, and
\item[{\bf (b)}] satisfies  a  $L$-Lipschitz
condition  \eqref{eq5.5}  for some
 positive bounded linear operator $L:B\to B$.
 \end{itemize}
 Then
\begin{itemize}
\item[{\bf i)}]
there exists a unique solution $u(t)$ of the problem
\eqref{eq4.6} such that ${\|u(t)\| \le1}$ for any $t\ge 0$;
\item[{\bf ii)}]
any two solutions $u(t)$ and $w(t)$  of problem \eqref{eq4.6} with initial
data in  the unit ball $U$ satisfy the inequality
\begin{equation}\label{eq5.11}
|u(t) - w(t)| \le \exp\{t(L-1)\} (|u_0-w_0|)\ .
\end{equation}
\end{itemize}
\end{theorem}

Note that under the same conditions as in  Theorem~\ref{theorem-lip}
 the operator $\Gamma$ given in Eqs.~\eqref{eq4.3} satisfies necessary
conditions for  the Theorem~\ref{theorem5.2}.

We remind to the reader that the initial value problem \eqref{eq4.6}
appeared as a generalization of the initial value problem
\eqref{eq3.2} for a
characteristic function  $\varphi (x,t)$, i.e., for the Fourier
transform  of a  probability measure (see
Eqs.~\eqref{eq3.1}, \eqref{eq2.1}).
It is important therefore to show that the solution $u(x,t)$ of the
problem \eqref{eq4.6} is a characteristic function for any $t>0$
provided this is so for $t=0$, which is addressed  in the
following statement.

\begin{lemma}\label{lemma5.3}
Let $U' \subset U\subset B$ be any closed convex subset of the unit
ball $U$ (i.e., $u = (1-\theta) u_1 + \theta u_2 \in U'$ for any
$u_{1,2} \in U'$ and $\theta\in [0,1]$). If $u_0 \in U'$ in
Eq.~\eqref{eq4.6} and $U$ is replaced by $U'$ in the condition~(1)
of Theorem~\ref{theorem5.2}, the theorem holds and $u(t) \in U'$ for
any $t\ge 0$.
\end{lemma}

\begin{remark} It is well-known (see, for example, the textbook
\cite{Fe}) that the
set $U'\subset U$ of Fourier transforms of probability measures
in $\real^d$ (Laplace transforms in the case of $\real_+$) is convex
and closed with respect to uniform convergence.
On the other hand, it is easy to verify that the inclusion $\Gamma(U')
\subset U'$, where $\Gamma$ is given in Eqs.~\eqref{eq4.3}, holds in
both cases of Fourier and Laplace transforms.
Hence, all results obtained for Eqs.~\eqref{eq4.2}, \eqref{eq4.3} can
be interpreted in terms of ``physical'' (positive and satisfying the
condition \eqref{eq2.1}) solutions of corresponding Boltzmann-like
equations with multi-linear structure of any order.

We also point out that all results of this section remain valid for
operators $\Gamma$ satisfying conditions \eqref{eq4.3},  with a
more general condition such as
\begin{equation}\label{eq.5.10.bis}
\sum_{n=1}^N \alpha_n \le 1\ ,\qquad \alpha_n \ge 0\ ,
\end{equation}
so that  $\Gamma(1) <1 $ and so the mass may not be conserved. The
only difference in this case is that the operator $L$  satisfying
conditions \eqref{eq5.2}, \eqref{eq5.3} is not a linearization of
$\Gamma(u)$ near the unity, but  nevertheless
Theorem~\ref{theorem-lip} remains true. The inequality
\eqref{eq.5.10.bis} is typical for Fourier (Laplace) transformed
Smoluchowski-type equations where the total number of particles is
decreasing in time (see \cite{MePe1, MePe2} for related work).
\end{remark}

\vskip.5in


In the next three  sections we study in more detail the solutions to
the initial value   problem \eqref{eq4.6}-\eqref{eq4.7} constructed
in Theorem~\ref{theorem5.2} and,  in particular, their long time
behavior, existence, uniqueness,
 and properties of the self-similar solutions.

\subsection{Large time asymptotics}\label{sec:5.2}

The long time asymptotics results
 are  a consequence of  some very general properties of
 operators
$\Gamma$,
 namely, that $\Gamma$ maps the unit ball $U$ of the
Banach space
$B=C(\real_+)$ into itself,   $\Gamma$ is an  $L$-Lipschitz
operator (i.e. satisfies \eqref{eq5.5})  and that $\Gamma$ is invariant
under dilations.

These three properties are sufficient to study self-similar solutions
and large time asymptotic behavior for the solution to the Cauchy problem
\eqref{eq4.6} in the unit ball $U$ of the
Banach space $C(\real_+)$.

\bigskip

\vbox{
{\bf Main properties of the operator
$\Gamma$:}  \hfill
{\sl
 \begin{itemize}
\item[{\bf a)}] $\Gamma$ maps the unit ball $U$ of the Banach space
$B=C(\real_+)$ into itself, that is
\begin{equation}\label{eq6.1bis}
\qquad \|\Gamma (u)\| \le1\ \text{ for any } \ \ u\in C(\real_+)
\ \ \text{ such that} \ \
\|u\|\le 1.
\end{equation}
 \item[{\bf b)}]$\Gamma$ is an  $L$-Lipschitz
operator (i.e. satisfies \eqref{eq5.5}) with $L$ from \eqref{eq5.2}, i.e.
 \begin{equation}\label{eq6.1}
\qquad \  |\Gamma (u_1) - \Gamma (u_2)|(x)
\le L(|u_1-u_2|)(x) = \int_0^\infty da K (a) | u_1 (ax) - u_2 (ax)|\ ,\qquad
\end{equation}
for $K(a)\geq 0$, for all $x\ge 0$ and for any two functions $u_{1,2} \in C(\real_+)$ such that
$\|u_{1,2}\| \le 1$.
\item[{\bf c)}] $\Gamma$ is invariant
under dilations:
\begin{equation}\label{eq6.2}
e^{\tau\D} \Gamma (u)  = \Gamma (e^{\tau\D} u)\ ,\quad
\D = x \frac{\partial}{\partial x}\ ,\quad
e^{\tau\D} u(x) = u(xe^\tau),\quad \tau\in\real\ .
\end{equation}
\end{itemize}
}
}

No specific information about $\Gamma$ beyond these three conditions
will be used in this section.

It was already shown in Theorem~\ref{theorem5.2} that the conditions
{\bf a)}  and {\bf b)} guarantee existence and uniqueness of the
solution $u(x,t)$ to the initial value
problem~\eqref{eq4.6}-\eqref{eq4.7}. The property {\bf b)} yields
the estimate \eqref{eq5.11} that is very important for large time
asymptotics, as we shall see below. The property   {\bf c)} suggests
a special class of self-similar solutions to Eq.~\eqref{eq4.6}.

We recall  the usual  meaning of the notation $y = O(x^p)$
(often used below): $y = O(x^p)$
 if and only if  there exists a  positive constant
$C$ such that
\begin{equation}\label{big-o}
 |y(x)| \le C x^p \ \ \ \   \text{for \ any} \ \ \
x\ge 0.
\end{equation}

In order to study long time stability properties to solutions whose
 initial data differs in terms of $ O(x^p) $, we will need some
 spectral properties of  the linear operator $L$.

\begin{definition}\label{spectra}
Let $L$ be the positive  linear operator
 given in Eqs.~\eqref{eq5.2}, \eqref{eq5.3},then
\begin{equation}\label{eq6.6}
L x^p = \lambda (p) x^p\ ,\qquad \ \ \ \
0< \lambda (p) = \int_0^\infty da\, K(a) a^p <\infty\ , \quad\ p\geq 0\, ,
\end{equation}
and the spectral function $\mu(p)$  is defined by
\begin{equation}\label{spec-func}
\mu(p) = \frac{\lambda (p)-1} p\ .
\end{equation}
\end{definition}

An immediate consequence of properties  {\bf a)} and
{\bf b)}, as stated in \eqref{eq6.1}, is that
one can obtain a criterion for a  point-wise in $x$
estimate  of   the difference of two solutions to
the initial value problem~\eqref{eq4.6}  yielding
 decay properties depending on the spectrum of $L$, as the following
statement and its corollary assert.

 \begin{lemma}\label{lemma6.2n}
 Let $u_{1,2}(x,t)$ be any  two classical solutions of the problem
 \eqref{eq4.6} with  initial data   satisfying the conditions
 \begin{equation}\label{eq6.7n}
|u_{1,2}(x,0)| \leq1, \ \ \ \  \ |u_1(x,0) - u_2(x,0)| \leq C\,x^p\, ,\ \ x\geq0
 \end{equation}
 for some positive constant $C$ and   $p$.
 Then
  \begin{equation}\label{eq6.8n}
 |u_1(x,t) - u_2 (x,t)| \le C{x^p}  \,  e^{-t(1-\lambda(p))} \, , \qquad\text{for\  all}\ \ t\geq 0
\end{equation}
 \end{lemma}

\begin{corollary}\label{cor00}
The minimal constant $C$ for which condition \eqref{eq6.7n}
is satisfied is
\begin{equation}\label{eq6.9n}
C_0= \sup_{x\geq 0} \frac{|u_1(x,0) - u_2(x,0)| }{x^p}=
\left\|\frac{u_1(x,0) - u_2(x,0)| }{x^p} \right\|\, ,
\end{equation}
and the following estimate holds
\begin{equation}\label{eq6.10n}
\left\|\frac{u_1(x,t) - u_2(x,t)| }{x^p} \right\| \leq
e^{-t(1-\lambda(p))}\left\|\frac{u_1(x,0) - u_2(x,0)| }{x^p} \right\|
\end{equation}
for any $p>0$.
\end{corollary}

A result similar to Lemma~\ref{lemma6.2n} was first obtained in
\cite{BC-03} for the inelastic Boltzmann equation whose Fourier
transform  is given in example {\bf (C)}, Eq.~\eqref{eq2.10}. Its
corollary in the form similar to \eqref{eq6.10n} for equation
~\eqref{eq2.10} was stated later in \cite{BCT-03} and was
interpreted there as ``the contraction property of the Boltzmann
operator'' (note that the left hand side of Eq.\eqref{eq6.10n} can
be understood as a distance between two solutions). Independently of
the terminology. the key reason for estimates
\eqref{eq6.8n}-\eqref{eq6.10n} is the Lipschitz property of the
operator $\Gamma$. It is remarkable that the large time asymptotics
of $u(x,t)$, satisfying the problem \eqref{eq4.6} with such
$\Gamma$, can be explicitly expressed through spectral
characteristics of the linear operator $L$.

In order to study the large time asymptotics of   $u(x,t)$ in more
detail we distinguish two different kinds of asymptotic behavior:
\begin{itemize}
\item[{\bf 1)}] convergence to stationary solutions,
\item[{\bf 2)}] convergence to self-similar solutions provided the
  condition ({\bf c}), of the main properties on $\Gamma$, is satisfied.
\end{itemize}

The case {\bf 1)} is relatively simple. Any stationary solution
$\bar u(x)$ of the problem  \eqref{eq4.6} satisfies the equation
 \begin{equation}\label{eq6.11n}
\Gamma(\bar u)=\bar u\, , \qquad  \bar u \in C(\real_+)\, , \ \
\|\bar u\|\leq 1\, .
 \end{equation}

If the stationary solution $\bar u(x)$  does exists (note, for example,
 that
$\Gamma(0)=0$ and $\Gamma(1)=1$ for $\Gamma$ given in
 Eqs.~\eqref{eq4.3})
then the large time asymptotics of some classes of initial data
$u_{0}(x)$ in \eqref{eq4.6} can be studied directly on the basis of
Lemma~\ref{lemma6.2n}.
 It is enough to assume that ${|u_0(x) - \bar u(x)| }$ satisfies
\eqref{eq6.7n}
 with $p$ such that $\lambda(p) <1$. Then $u(x,t) \to \bar u(x)$ as
 $t\to\infty$, for any $x\geq0$.

This simple consideration, however, does not answer at least two questions:
\begin{itemize}
\item[{\bf A)}] What happens with $u(x,t)$ if the inequality~\eqref{eq6.7n}
 for  ${|u_0(x) - \bar u(x)| }$ is satisfied with such $p$  that
$\lambda(p)>1$?
\item[{\bf B)}] What happens with $u(x,t)$ for large $x$
 (note that the estimate \eqref{eq6.8n} becomes trivial if $x\to\infty$).
\end{itemize}

In order to address these questions we consider a special class of solutions
 of Eq.~\eqref{eq4.6}, the so-called self-similar solutions. Indeed the
 property {\bf c)} of $\Gamma$ shows that  Eq.~\eqref{eq4.6} admits a
class of formal solutions $u_s(x,t)=w(x\, e^{{\mu_*}t}) $ with
some real $\mu_*$.
It is convenient for our goals to use a terminology that slightly differs
 from
 the usual one.

\begin{definition}\label{self-similar} The function
 $w(x)$ is called a  self-similar solution associated with the initial
 value problem~\eqref{eq4.6}
if it satisfies the problem
\begin{equation}\label{eq6.8}
\mu_* \D w + w = \Gamma (w)\ ,\qquad \|w\| \le 1\ ,
\end{equation}
in the notation of Eqs.~\eqref{eq6.2}, \eqref{eq4.3}.
\end{definition}

The convergence of solutions $u(x,t)$ of the initial value
problem~\eqref{eq4.6}
to a stationary solution $\bar u(x)$
can be considered as a special case of the self-similar asymptotics
with $\mu_*=0$.

 Under the assumption that self-similar solutions exists
(the existence is proved in the next section), we state the fundamental
result on the convergence  of solutions $u(x,t)$ of the initial value
problem~\eqref{eq4.6}  to self-similar ones (sometimes called in the
 literature {\sl  self-similar stability}).

\begin{lemma}\label{lemma6.1}
We assume that
\begin{itemize}
\item[{\bf i)}] for some $\mu_*\in\real$, there exists a classical
(continuously
differentiable if $\mu_*\ne0$) solution $w(x)$ of Eq.~\eqref{eq6.8} such
that $\|w\|\le 1$;
\item[{\bf ii) }] the initial data $u(x,0)= u_0$ in the problem \eqref{eq4.6}
 satisfies
\begin{equation}\label{eq6.9}
u_0 = w + O(x^p)\ ,\qquad \|u_0\| \le 1\ ,\text{ for } \ p>0\ \text{ such that }
\mu (p) <\mu_* ,
\end{equation}
where $\mu(p)$  defined in
 \eqref{spec-func} is  the spectral function
 associated to the operator $L$.
\end{itemize}
Then
\begin{equation}\label{eq6.12.1}
|u(xe^{-\mu_* t},t) - w(x)| = O({x^p}) e^{-pt(\mu_* -\mu(p))} \,
\end{equation}
 and therefore
\begin{equation}\label{eq6.12}
\lim_{t\to\infty} u(xe^{-\mu{_*} t}, t) = w(x)\ ,\qquad x\ge 0\ .
\end{equation}
\end{lemma}

\begin{remark} Lemma~\ref{lemma6.1}
 shows how to find a domain of attraction of any
self-similar solution provided the self-similar
 solution is itself  known.
It is remarkable that the domain of attraction can be expressed
in terms of just  the {\sl spectral function}  $\mu(p),\ p>0$,
defined in~\eqref{spec-func}, associated with
the linear operator $L$ for which the operator $\Gamma$ satisfies
the $L$-Lipschitz condition.

Generally speaking, the equality  \eqref{eq6.12} can be also
fulfilled for some other values of $p$ with $\mu (p) >\mu_*$
 in Eq.~\eqref{eq6.9},
but, at least, it always holds if $\mu (p)<\mu_*$.
\end{remark}

\medskip

We shall need some  properties of the
{\sl spectral function}  $\mu (p)$.
Having in mind further applications, we formulate these properties in terms
of the operator $\Gamma$ given in Eqs.~\eqref{eq4.3}, though they depend only
 on $K(a)$ in Eqs.~\eqref{eq6.6}

\begin{lemma}\label{lemma6.2}
The spectral function $\mu (p)$ has the following properties:
\begin{itemize}
\item[{\bf i)} ] It is positive and unbounded as $p\to 0^+$, with
 asymptotic behavior given by
\begin{equation}\label{eq6.13}
\mu (p) \approx \frac{\lambda (0) -1}p\ ,\qquad
p\to 0\ ,
\end{equation}
where, for $\Gamma$ from \eqref{eq4.3}
\begin{equation}\label{eq6.14}
\lambda(0) = \int_0^\infty da\, K (a)
= \sum_{n=1}^N \alpha_n n\ge 1\ ,\qquad
\sum_{n=1}^N \alpha_n =1\ ,\ \alpha_n \ge 0\ ,
\end{equation}
and therefore $\lambda (0)=1$ if and only if the operator  $\Gamma$
\eqref{eq4.3} is linear $(N=1)$;
\item[{\bf ii)}] In the case of a multi-linear $\Gamma$ operator,
 there is not more than one point $0<p_0 <\infty$, where
the spectral function $\mu(p)$ achieves its minimum, that is,
$\mu'(p_0)=\frac{d\,\mu}{d\,p} (p_0) =0$, with  $\mu (p_0) \le \mu (p)$ for any $p>0$,
provided  $N\ge 2$ and $\alpha_N >0$.
\end{itemize}
\end{lemma}

\begin{remark} From now on, we
shall always assume below that the operator $\Gamma$ from~\eqref{eq4.3}
is multi-linear.
Otherwise it is easy to see that the problem \eqref{eq6.8} has no
solutions (the condition $\|w\| \le 1$ is important!) except for  the trivial
ones $w=0,1$.
\end{remark}

The following corollaries are readily obtained from
lemma~\ref{lemma6.2} part {\bf ii)} and its proof.

\begin{corollary}\label{cor0}
For the case of a non-linear $\Gamma $ operator, i.e. $N\geq 2$,
the spectral function $\mu(p)$ is always monotone decreasing in the interval
$(0,p_0)$, and $\mu(p) \geq \mu(p_0)$ for $0 <p<p_0$. This implies that
 there exists
a unique inverse function
 $\tp(\mu):(\mu(p_0), +\infty) \to (0,p_0) $, monotone decreasing
in its domain of definition.
\end{corollary}

\begin{corollary}\label{cor1}
There are precisely four different kinds of qualitative behavior of
$\mu (p)$ shown on Fig.1.
\end{corollary}

\begin{proof}
There are two options: $\mu (p)$ is monotone decreasing function
(Fig.1~(a)) or $\mu (p)$ has a minimum at $p=p_0$ (Fig.1~{(b,c,d)}).
In  case  Fig.1~{(a)} $\mu (p) >0$ for all $p>0$ since $\mu (p) > 1/p$.
The asymptotics of $\lambda (p)$ \eqref{eq6.6} is clear:
\begin{align} 
   \text{{\bf (1)}}\qquad &\lambda (p) \xrightarrow[p\to\infty]{}
\lambda_\infty
   \in \real_+\ \text{ if }\  \int_{1^+}^\infty da\, K(a) = 0\ ;
\label{eq6.16} \\
   \text{{\bf (2)}}\qquad &\lambda (p) \xrightarrow[p\to\infty]{}\
\infty \text{ if }\
   \int_{1^+}^\infty da\, K(a) >0\ .
\end{align}

In the case {\bf (1)} when $\mu (p) \to\infty$ as $p\to 0$, two possible
pictures (with and without minimum) are shown on Fig.1~{(b)} and
Fig.1~{(a)} respectively.
In case {\bf (2)}, from Eq.~\eqref{eq6.6} it is clear that
$\lambda (p)$ grows exponentially for large $p$,
therefore $\mu (p) \to \infty$ as $p\to \infty$.
Then the minimum always exists and we can distinguish two cases:
$\mu (p_0) <0$ (Fig.1~{(d}) and $\mu (p_0) >0$ (Fig.1~{(c)})
\end{proof}
\begin{figure}
 \centering
\label{figure1}
\includegraphics[scale=.28]{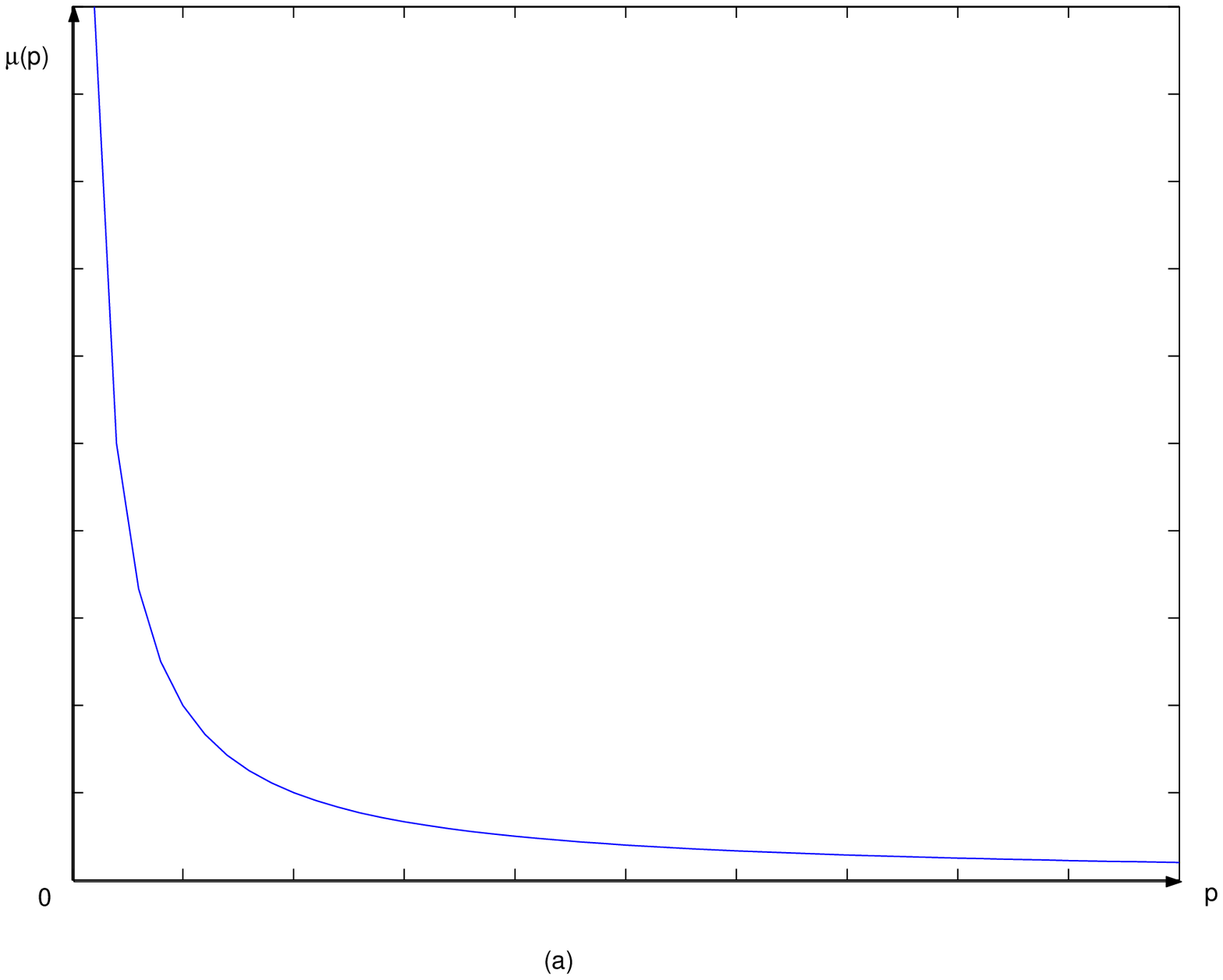}\qquad
\includegraphics[scale=.28]{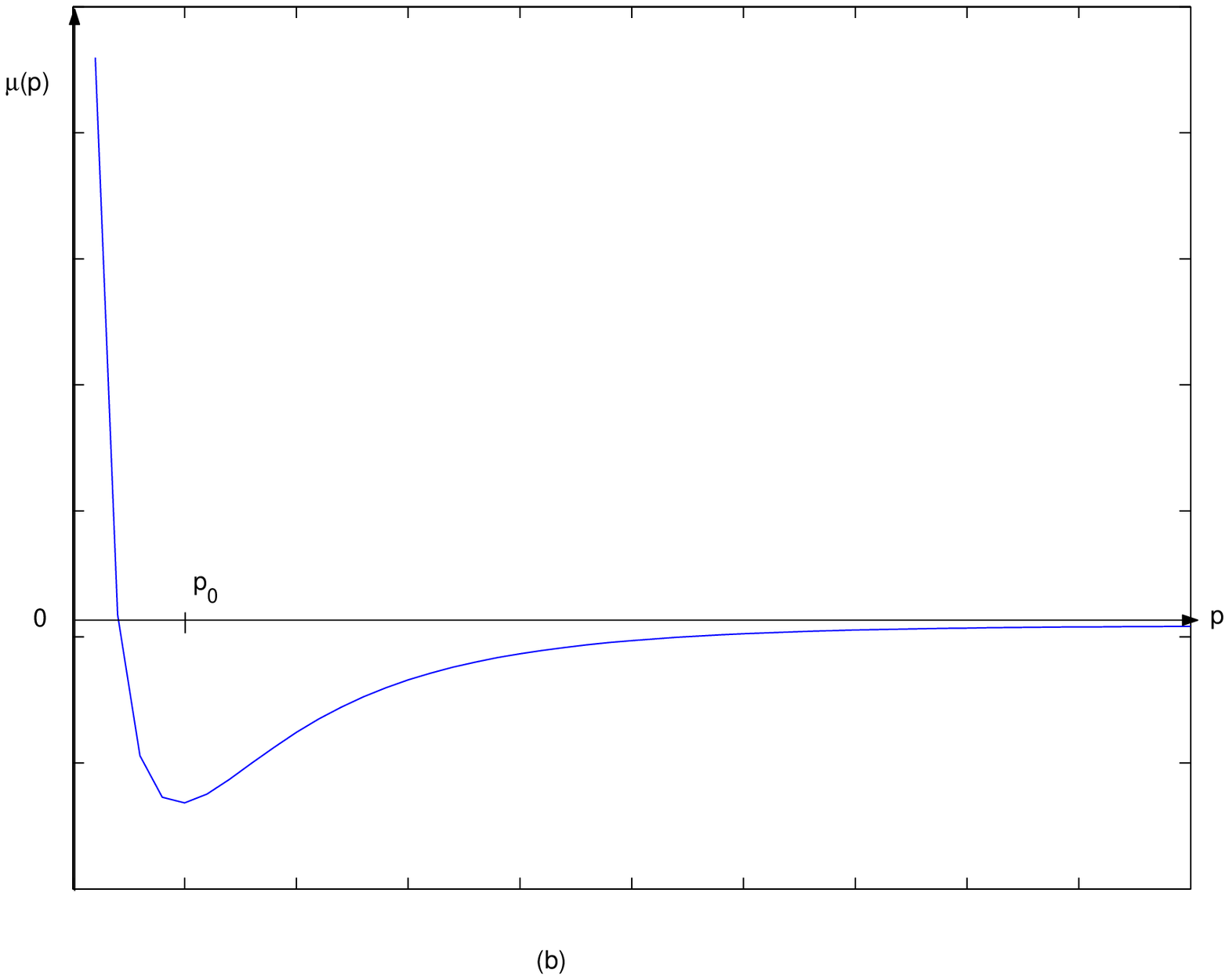}\\
\ \\
\includegraphics[scale=.28]{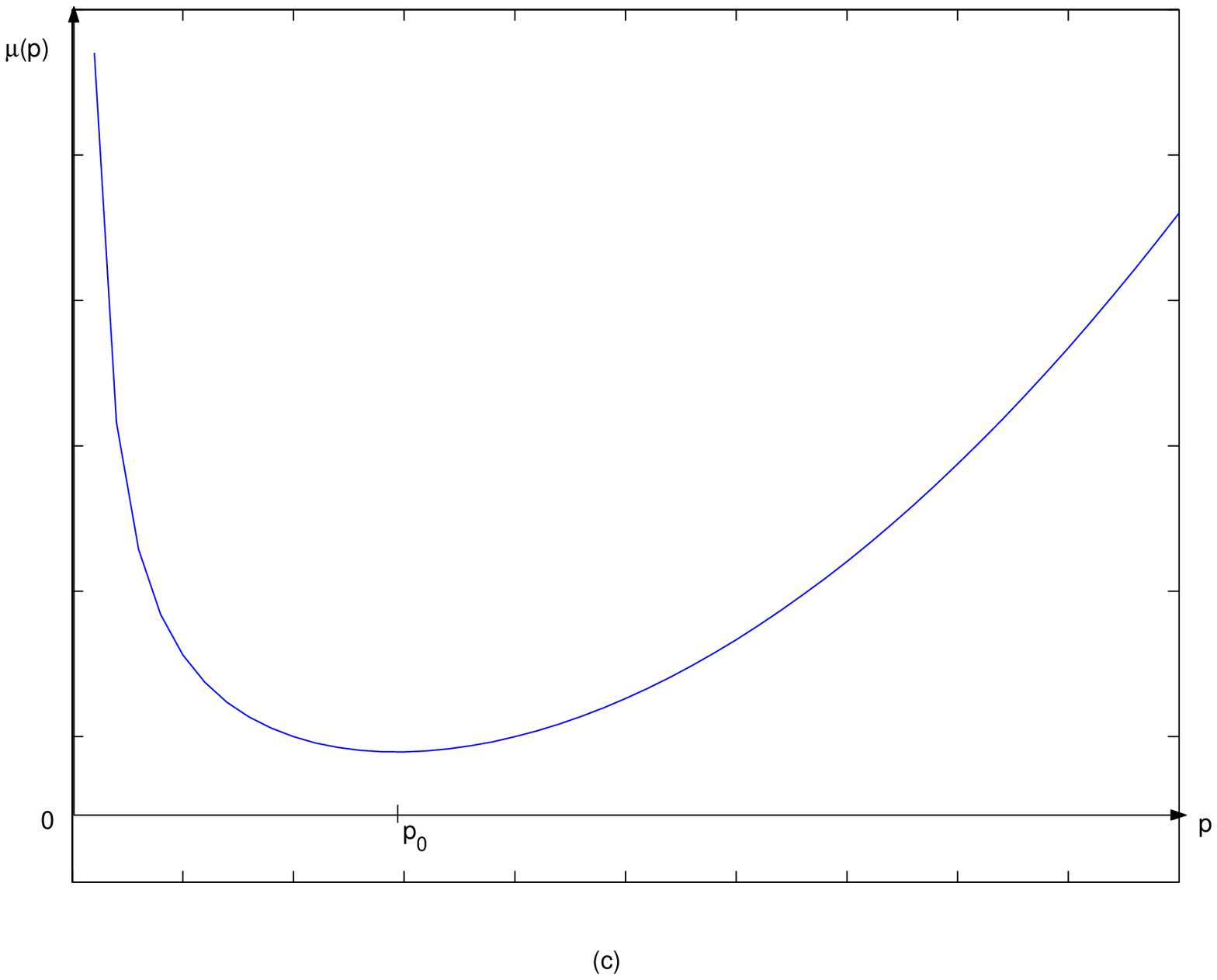}\qquad
\includegraphics[scale=.28]{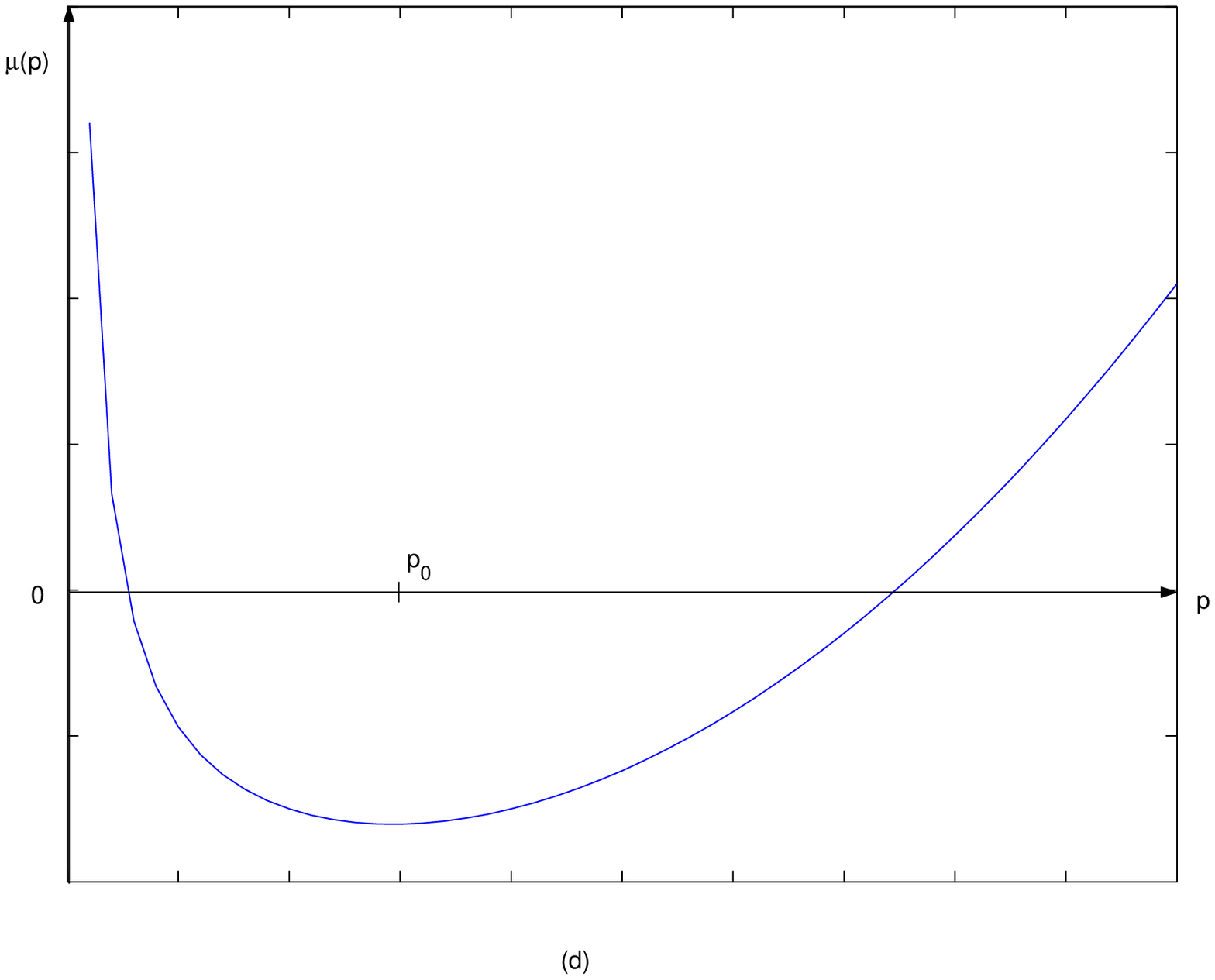}
\caption{Possible profiles of the spectral function $\mu(p)$ }
\end{figure}

We note that, for Maxwell models {\bf (A), (B), (C)} of Boltzmann
equation (Sections~2 and 3), only cases {(a)} and {(b)} of Fig.1 can
be possible (actually this is the case {(b)}) since the condition
\eqref{eq6.16} holds. Fig.1 gives a clear graphic representation of
the domains of attraction of self-similar solutions
(Lemma~\ref{lemma6.1}): it is sufficient to draw the line $\mu (p) =
\mu_* = {constant}$, and to consider a $p$ such  that the graph of
$\mu (p)$ lies below this line.

Therefore, the following corollary follows directly from the properties
of the spectral function $\mu(p)$, as characterized by the  behaviors in
 Fig.1, where
we assume that
 $\mu (p_0)=0$ for $p_0= \infty$, for the case shown on Fig.1~{(a)}.

\begin{corollary}\label{cor2}
Any self-similar solution $u_s(x,t) = w(xe^{\mu_* t})$ with $\mu (p_0)
<\mu_* <\infty$ has a non-empty domain of attraction, where $p_0$ is the unique
(minimum) critical point of the spectral function $\mu(p)$.
\end{corollary}

\begin{proof}
We use Lemma~\ref{lemma6.1} part {\bf ii)} on any initial state
$u_0= w + O(x^p)$ with $p>0$ such that $\mu (p_0) \leq \mu (p)
<\mu_* $. In particular, Eqs.~\eqref{eq6.12.1} and   \eqref{eq6.12}
show that the domain of attraction of  $ w(xe^{\mu_* t})$ contains
any
 solution to the initial value problem~\eqref{eq4.6} with the initial
state as above.
\end{proof}

\medskip

It is clear that  the inequalities of the kind
$u_1 - u_2 = O (x^p)$
for\  any $p>0 $ such  that
$\mu (p) <\mu_*$,   for any fixed
$\mu_* \geq \mu(p_0)$
play an important role.
We can use specific properties of $\mu (p)$ in order to express such
inequalities in more convenient form.

\begin{lemma}\label{lemma6.3}
 For any given $\mu_* \in (\mu (p_0),\infty)$ and
$u_{1,2}(x)$ such that $\|u_{1,2}\| <\infty$,
 the following two statements are equivalent:
\begin{itemize}
\item[{\bf i)}]
there exists  $p>0$ such that
\begin{equation}\label{eq6.18}
u_1 - u_2 = O (x^p)\ ,\qquad  \text{with}\ \
\mu (p) <\mu_*\, .
\end{equation}

\item[{\bf ii)}] There exists $\ep >0$ such that
\begin{equation}\label{eq6.19}
u_1 - u_2 = O(x^{{\tp}(\mu_*) +\ep})\ ,
\quad \ \text{ with}\ \
 \tp(\mu_*) < p_0\ ,
\end{equation}
where $\tp(\mu)$ is the inverse to $\mu (p)$ function, as defined
in {\bf Corollary~\ref{cor0}}.
\end{itemize}
\end{lemma}

\medskip

Finally, to conclude this section,
 we show  a general property of the initial value  problem~\eqref{eq4.6}
for any non-linear $\Gamma$ operator
 satisfying conditions {\bf a)} and {\bf b)}
given in Eqs.~\eqref{eq6.1bis} and \eqref{eq6.1} respectively.
This property gives the control to the point-wise difference
of any  two rescaled
 solutions to \eqref{eq4.6} in the unit sphere of B , whose
initial states differ by $O(x^p)$. It is formulated as follows.

\begin{lemma}\label{lemma6.4}
Consider the problem \eqref{eq4.6}, where $\Gamma$ satisfies the
conditions {\bf (a)} and {\bf (b)}.
Let $u_{1,2} (x,t)$ are two solutions satisfying the initial
conditions $u_{1,2} (x,0) = u_0^{1,2} (x)$ such that
\begin{equation}\label{eq6.20}
\|u_0^{1,2}\| \le 1\ ,\quad
u_0^{1} - u_0^{2} = O(x^p)\ ,\quad
p>0\ .
\end{equation}
then, for any  real $\mu_*$,
\begin{equation}\label{eq6.21}
\Delta_{\mu_{*}} (x,t) = u_{1} (xe^{-\mu_* t},t) - u_{2} (xe^{-\mu_* t},t)
= O( x^p) e^{-pt[\mu_* -\mu (p)]}
\end{equation}
and therefore
\begin{equation}\label{eq6.22}
\lim_{t\to \infty} \Delta_{\mu_{*}} (x,t) = 0\ ,\qquad x\ge0\ ,
\end{equation}
for any $\mu_* >\mu (p)$.
\end{lemma}

\begin{remark} There  is an important point to understand here.
Lemmas~\ref{lemma6.2n} and~\ref{lemma6.1}  hold for any operator
$\Gamma$ that satisfies just the  two properties {\bf (a) }and
{\bf(b)} stated in \eqref{eq6.1bis} and \eqref{eq6.1}. It says that,
in some sense, a distance between any two solutions with initial
conditions satisfying Eqs.~\eqref{eq6.20} tends to zero as $t\to
\infty$. Such terminology and corresponding distances were
introduced for specific forms of Maxwell-Boltzmann models in
\cite{GTW,BCT-06}. It should be pointed out, however, that this {\sl
contraction  property} may not say much about large time asymptotics
of $u(x,t)$, unless the corresponding self-similar solutions are
known, for which the operator $\Gamma$ must be invariant under
dilations (so it satisfies also
 property {\bf (c)} as well, as stated in \eqref{eq6.2} ).
In such case
one can use estimate~\eqref{eq6.22} to deduce the convergence
in the form \eqref{eq6.12},  \eqref{eq6.13}.
\end{remark}

Therefore one must study  the problem of existence of self-similar
 solutions, which is  considered in the next
Section.

\vskip.5in

\subsection{Existence of self-similar solutions}\label{sec:5.3}

We develop now a criteria for existence, uniqueness
and self-similar asymptotics to the    problem  Eq.~\eqref{eq6.8} for any
operator $\Gamma$  that satisfies conditions  {\bf (a)},
{\bf (b)} and {\bf (c)} from Section~6, with the
 corresponding spectral function $\mu(p)$ defined in
\eqref{spec-func}.

Theorem~\ref{theorem7.1} below shows the criteria for
 existence and uniqueness of
self-similar solutions for any operator $\Gamma$ that satisfies just
conditions  {\bf (a)} and {\bf (b)}.  Then  Theorem~\ref{theorem7.2}
follows,  showing a general criteria to self-similar asymptotics for
the problem \eqref{eq4.6} for any operator $\Gamma$  that satisfying
conditions  {\bf (a)} {\bf (b)} and {\bf (c)}.

We consider Eq.~\eqref{eq6.8} written in the form
\begin{equation}\label{eq7.1}
\mu xw' (x) + w(x) = g(x)\ ,\qquad
g = \Gamma (w)\ ,\ \mu \in \real\ ,
\end{equation}
and, assuming that $\|w\| <\infty$, transform this equation to the
integral form.
It is easy to verify that the resulting integral equation reads
\begin{equation}\label{eq7.2}
w(x) = \int_0^1 d\tau\, g(x\tau^\mu)
\end{equation}

By means of an iteration scheme, the following result can be proved.
\begin{theorem}\label{theorem7.1}
Consider Eq.~\eqref{eq7.1} with arbitrary $\mu\in\real$ and
the operator $\Gamma$ satisfying the conditions {\bf (a) }and
{\bf (b) }
from Section~6.
Assume that there exists a continuous function $w_0(x)$, $x\ge0$,
such that
\begin{itemize}
\item[{\bf  i)}] \ \ \ \ \ \ \ \ \ $\|w_0\| \le 1$ and

\item[{\bf  ii)}]
\begin{equation}\label{eq7.10}
 \int_0^1 d\tau\, g_0 (x\tau^\mu) = w_0 (x) + O(x^p)\ ,\qquad
g_0 = \Gamma (w_0)\ ,\hfill
\end{equation}
\newline
with some $p>0$ satisfying the inequality
\begin{equation}\label{eq7.11}
\mu (p) = \frac1p [ \int_0^\infty da\, K(a) a^p -1] <\mu\ .
\end{equation}
\end{itemize}
Then there exists a classical solution $w(x)$ of Eq.~\eqref{eq7.1}.
The solution is unique in the class of continuous functions
satisfying conditions
\begin{equation}\label{eq7.12}
\|w\| \le 1\ ,\qquad
w(x) = w_0 (x) + O(x^{p_1})\ ,
\end{equation}
with any $p_1$ such that $\mu (p_1) <\mu$.
\end{theorem}

 Combining Lemma~\ref{lemma5.1}
 and  Lemma~\ref{lemma6.1},
  the following  general statement related to the self-similar asymptotics
 for the problem \eqref{eq4.6} is obtained.

 \begin{theorem}\label{theorem7.2}
 Let $u(x,t)$ be a solution of the problem \eqref{eq4.6} with $\|u_0\|\le1$
 and $\Gamma$ satisfying the conditions {\bf (a)},
 {\bf (b)}, {\bf (c)} from Section~6.
 Let $\mu (p)$ denote the spectral function \eqref{eq7.11} having its
 minimum (infimum) at $p = p_0$ (see Fig.1), the case $p_0 = \infty$
 is also included.
 We assume that there exists $p \in (0,p_0)$ and $0<\ep<p_0-p$ such that
 \begin{equation}\label{eq7.13a}
 \int_0^1 d\tau\, g_0 (x\tau^{\mu (p)})
 = u_0 (x) + 0 (x^{p+\ep})\ ,\qquad
 g_0 = \Gamma (u_0),\quad  \ep>0\ .
 \end{equation}
 Then
\begin{itemize}
\item[{\bf i)}] there exists a unique solution $w(x)$ of the
equation~\eqref{eq7.1} with $\mu = \mu (p)$  such that
\begin{equation}\label{eq7.14}
\|w\| \le 1\ ,\qquad w(x) = u_0 (x) + O (x^{p+\ep})\ ,
\end{equation}
\item[{\bf ii)}]
\begin{equation}\label{eq7.15}
\lim_{t\to\infty} u (x\,e^{-\mu (p)t} ,t) = w(x)\ ,\qquad x\ge 0\ ,
\end{equation}
where the convergence is uniform on any bounded interval
in $\real_+$ and
\begin{equation}\label{eq7.17}
u (x\, e^{-\mu (p)t} ,t) -  w(x)= O(x^{p+\ep}e^{-\beta(p,\ep)t} )\, ,
\end{equation}
with $\beta(p,\ep)= (p+\ep)(\, \mu(p) -\mu(p+\ep)\, ) >0$.
\end{itemize}
\end{theorem}

Hence,  a general criterion \eqref{eq7.14} is obtained
for  the self-similar
asymptotics of $u(x,t)$ with a  given initial condition $u_0 (x)$.
The criterion can be applied to the problem \eqref{eq4.6} with any
operator $\Gamma$ satisfying conditions
{\bf (a)}, {\bf (b)}, {\bf (c)} from Section~6.
The specific class \eqref{eq4.3} of operators $\Gamma$ is studied in
Section~8. We shall see below that the condition~\eqref{eq7.14}
can be essentially simplified for such operators.


 \subsection{Properties of self-similar solutions}\label{sec:5.4}

We now apply the general theory (in particular,
Theorem~\ref{theorem7.2}) to the particular case of the multi-linear
 operators $\Gamma$ considered in Section 4, where their
corresponding  spectral function $\mu (p)$
satisfies \eqref{eq7.11}, \eqref{eq5.3} whose behavior corresponds to
 Fig.1.  We also show that  $p_0= \min_{p>0} \mu(p) >1$ is a necessary
 condition for self-similar asymptotics.

In addition, Theorem~\ref{theorem8.3} establish sufficient
conditions for which self-similar solutions of problem \eqref{eq7.1}
will lead to well defined
 self-similar solutions (distribution functions) of the
original problem after taking the inverse Fourier transform.

\medskip

 We consider the integral equation \eqref{eq7.2} written as
 \begin{equation}\label{eq8.1}
 w= \Gamma_\mu (w) = \int_0^1 dt\, g(xt^\mu)\ ,\qquad
 g = \Gamma (w)\ ,\ \mu \in\real\ .
 \end{equation}
 The following  two properties of $w(x)$ that are independent of
the  specific form \eqref{eq4.3} of $\Gamma$.

 \begin{lemma}\label{lemma8.1}
 $\quad$
 \begin{itemize}
 \item[{\bf i.}]  If there exist a closed subset $U' \subset U$ of the unit ball
 $U$  in $B$,  such that
$\Gamma _\mu (U')\subset U'$
 for any $\mu \in\real$, and for some function $w_0 \in U'$ the conditions
 of Theorem~\ref{theorem7.1} are satisfied,  then $w\in U'$.
\item[{\bf ii.}]  If the conditions of Theorem~\ref{theorem7.1} for $\Gamma$ are
 satisfied and, in addition, $\Gamma (1) =1$,
 \end{itemize}
 then the solution $w_* =1$
 of Eq.~\eqref{eq8.1} is unique in the class of functions $w(x)$
 satisfying the condition
 \begin{equation}\label{eq8.2}
 w(x) =1 + O (x^p)\ ,\qquad \mu (p) <\mu\ .
 \end{equation}
 \end{lemma}

We observe that the statement [ii] can be interpreted as a necessary
condition for existence of non-trivial ($w\ne \text{const.}$)
solutions of Eq.~\eqref{eq8.1}: if there exists a non-trivial
solution $w(x)$ of Eq.~\eqref{eq8.1}, where $\Gamma (1) =1$, such
that
\begin{equation}\label{eq8.3}
\|w\| =1\ ,\qquad w= 1 + O (x^p)\ ,\quad p>0\, ,\ \
\ \ \ \text{then} \ \ \ \mu \le \mu(p)\, .
\end{equation}

We recall that $\mu (p)$ satisfies the inequality $\mu (p) \ge \mu(p_0)$
(Fig.1).

If $p \ge p_0$ (provided $p_0 <\infty$) in Eqs.~\eqref{eq8.3}, then there
are  no non-trivial solutions with $\mu >\mu (p_0)$.

On the other hand, possible solutions with $\mu \le \mu(p_0)$ (even
if they exist) are irrelevant for the problem~\eqref{eq4.6} since
they have an  empty domain of attraction (Lemma~\ref{lemma6.1}).

Therefore we always assume below that $\mu >\mu (p_0)$ and,
consequently, $p \in (0,p_0)$ in Eq.~\eqref{eq8.3}.

\medskip

Let us consider now the specific class \eqref{eq4.3}--\eqref{eq4.3bis}
 of operators $\Gamma$,
with functions $u (x)$ satisfying the condition $u(0)=1$.
Then, $u(0,t) =1$ for the solution $u(x,t)$ of the
problem \eqref{eq4.6}.

In addition, the operators \eqref{eq4.3} are invariant under dilation
 transformations  \eqref{eq6.2} (property {\bf (c)}, Section 6).
Therefore, the problem \eqref{eq4.6} with the initial condition $u_0(x)$
satisfying
\begin{equation}\label{eq8.4}
u(0) =1\ ,\quad \|u_0\| = 1\ ;\qquad u_0 (x) =1 - \beta x^p + \cdots,\quad
x\to 0\ ,
\end{equation}
can be always reduced to the case $\beta =1$ by transformation
$x' = x\beta^{1/p}$.

Moreover, the whole class of operators \eqref{eq4.3} with different
kernels $A_n (a_1,\ldots,a_n)$, $n = 1,2,\ldots$, is invariant under
transformations $\tilde x = x^p$, $p>0$.
The result of such transformation of $\Gamma$ is another operator
$\widetilde\Gamma$ of the same class \eqref{eq4.3} with kernels
$\tilde A_n (a_1,\ldots,a_n)$.

Therefore we fix the initial condition \eqref{eq8.4} with $\beta =1$
and transform the function \eqref{eq8.4} and the equation \eqref{eq4.6}
to new variables $\tilde x = x^p$.
Then we omit the tildes and reduce the problem \eqref{eq4.6},
with initial condition
\eqref{eq8.4} to the case $\beta =1$, $p=1$.
We study this case in detail and formulate afterward the results
in terms of initial variables.

\medskip

Next, we assume a bit more about the asymptotics of the initial data
$u_0(x)$ for small  $x$, namely
\begin{equation}\label{eq8.5}
\|u_0\| =1\ ,\quad u_0 (x) = 1-x + O(x^{1+\ep})\ ,\quad x\to 0\ ,
\end{equation}
with some $\ep>0$.

Then, our goal now is to apply the general theory (in particular,
Theorem~\ref{theorem7.2} and  criterion \eqref{eq7.13a}) to this
particular case. We assume that the spectral function $\mu (p)$
given by \eqref{eq7.11}, \eqref{eq5.3},  corresponds to one of the
four cases shown on Fig.1 with $p_0 >1$.

Let us take a typical function $u_0 = e^{-x}$
satisfying \eqref{eq8.5}  and apply the criterion \eqref{eq7.13a},
 from Theorem~\ref{theorem7.2} or, equivalently, look for such   $p>0$   that
 \eqref{eq7.13a} is satisfied. That is, find possibles values of $p>0$
such that
\begin{equation}\label{eq8.6}
\Gamma_{\mu (p)} (e^{-x}) - e^{-x} = 0 (x^{p+\ep})\, ,
\end{equation}
in the notation of Eq.~\eqref{eq8.1}.

It is important to observe that now the spectral function $\mu (p)$
is closely connected with the operator $\Gamma$ (see
Eqs.~\eqref{eq7.11} and \eqref{eq5.3}), since this was not assumed
in the general theory of Sections~4--7. This connection leads to
much more specific results, than, for example, the general
Theorems~\ref{theorem7.1}, \ref{theorem7.2}.

The properties of self-similar solutions
to problem \eqref{eq7.1} for $p_0>1$, and consequently, for
\begin{equation}\label{eq8.9}
\mu (p) \ge \mu (p_0) > - \frac1{p_0} > -1\, ,
\end{equation}
can be obtained from the structure of
 $\Gamma_\mu (e^{-x})$ for any $\mu>-1$
using its  explicit formula  (Eqs.~\eqref{eq7.11}
and \eqref{eq5.3}). In particular, for
\begin{equation}\label{eq8.7}
\Gamma_\mu (e^{-x} )
= \sum_{n=1}^N \alpha_n \int_{\real_+^n} da_1\ldots da_n A_n
(a_1,\ldots,a_n) I_\mu \Big[ x \sum_{k=1}^n a_k\Big]\ ,
\end{equation}
where
\begin{equation}\label{eq8.8}
I_\mu (y) = \int_0^1 dt\, e^{-yt^\mu}\ ,\ \
\quad \mu\in\real\, , \  y>0\, ,\qquad \qquad
\sum_{n=1}^N \alpha_n =1\, .
\end{equation}
 the following two
statements can be proven.

    \begin{lemma}\label{cor6}
 The condition \eqref{eq8.6} is fulfilled if and only if
 $p\le 1$, and therefore $\mu (p) \ge \mu (1)$ whenever
$p_0 > 1$ with $\mu(p_0)=\min_{_{p>0}}\mu(p)$.
\end{lemma}

 \begin{theorem}\label{theorem8.3}
 The limiting function $w(x)$ constructed by
 the iteration scheme to prove existence in
the solution $w$ in Theorem~\ref{theorem7.1}
 satisfies Eq.~\eqref{eq8.1}
 with $\mu = \mu (1)$,
where $\Gamma$ is
 given in Eqs.~\eqref{eq4.3}, $\mu (p)$ is defined in Eqs.~\eqref{eq7.11},
\eqref{eq5.3}. Then, the following  conditions  are
 fulfilled for $w(x)$:
\begin{itemize}
\item[{\bf 1.}] It satisfies
 \begin{equation}\label{eq8.19}
 0\le < w(x) \le 1\ ,\qquad \text {with} w(0) =1\  \ \text{and}\ \ w'(0) = -1\ ,
 \end{equation}
 \begin{equation}\label{eq8.20}
 w'(x) \le 0\ ,\qquad |w'(x)| \le 1\ , \qquad  \text{and }\ \ \ w(x) = e^{-x} + O (x^{\pi (\mu)})\ ,
 \end{equation}
 with
\begin{displaymath}\label{eq8.22}
\pi (\mu) = \left\{ \begin{array}{ll}
2 & \textrm{if $\mu >-\frac12,$}\\
2-\ep \quad\text{for any }\ep> 0 \ \ \  & \textrm{if $\mu=\frac12, $ }\\
\frac1{|\mu|}  & \textrm{if $-1 <\mu < -\frac12.$ }
\end{array} \right.
\end{displaymath}
\item[{\bf 2.}] Further
 \begin{equation}\label{eq8.23}
 e^{-x}\le  w (x) \le 1 \ ,\qquad
 \lim_{x\to\infty} w(x) = 0\ , \qquad \text{and}
 \end{equation}
\item[{\bf 3.}]  there exists a generalized non-negative function
 $R(\tau)$, $\tau \ge 0$, such that
 \begin{equation}\label{eq8.24}
 w(x) = \int_0^\infty d\tau\, R(\tau) e^{-\tau x}\ ,\qquad
 \int_0^\infty d\tau\, R(\tau)
 = \int_0^\infty d\tau\, R(\tau) \tau =1\ .
 \end{equation}
\end{itemize}
 \end{theorem}

The integral representation \eqref{eq8.24} is important for
properties of corresponding distribution functions satisfying
Boltzmann-type equations.
Now it is easy to return to initial variables with $u_0$
given in Eq.~\eqref{eq8.4} and to describe the complete
picture of the self-similar relaxation for the problem \eqref{eq4.6}.


 \section{Main results for Maxwell models
 with multiple interactions}\label{sec:6}

\subsection{Self-similar asymptotics}\label{sec:6.1}

We apply now  the results of Section~6 to the specific case when
  the Cauchy problem \eqref{eq4.6} with a fixed operator
  $\Gamma$ \eqref{eq4.3} corresponds to the Fourier transform problem for
 Maxwell models
 with multiple interactions. In particular we
 study the time evolution of $u_0(x)$
  satisfying the conditions
  \begin{equation}\label{eq9.1}
  \|u_0\| = 1\ ; \quad u_0 = 1 - x^p + O(x^{p+\ep})\ ,\   x\to 0\ ,
  \end{equation}
  with some positive $p$ and $\ep$.
  Then, from Theorems~\ref{theorem7.1} and \ref{theorem7.2},
  there exists a unique classical solution $u(x,t)$ of the
  problem \eqref{eq4.6}, \eqref{eq9.1} such that, for all $t\ge0$,
  \begin{equation}\label{eq9.2}
  \|u(\cdot,t) \| =1\ ;\qquad
  u(x,t) =1 + O(x^p)\ ,\ x\to 0\ .
  \end{equation}

First, consider the linearized operator $L$ given in
  Eqs.~\eqref{eq5.2}--\eqref{eq5.3} and construct the spectral
  function $\mu (p)$ given in Eq.~\eqref{eq7.11} which
 will be of one of four kinds
  described qualitatively on Fig.1.

Second,
  find the value $p_0 >0$ that corresponds to minimum (infimum)
  of $\mu (p)$.
  Note that $p_0 =\infty$ just for the case described on Fig.1~(a),
  otherwise $0<p_0<\infty$.
  Compare $p_0$ with the value $p$ from Eqs.~\eqref{eq9.1}.
  If $p<p_0$ then the problem \eqref{eq4.6}, \eqref{eq9.1} has a
  self-similar asymptotics (see below).

 In particular, two different cases are
  possible: (1)~$p\ge p_0$ provided $p_0<\infty$;
  (2)~$0<p< p_0$.
  In the first case a behavior of $u(x,t)$ for large $t$ may depend
  strictly on initial conditions.

Depending on how $p$ compares with $p_0$, we can obtain
 We again use Lemma~\ref{lemma6.4} with $u_{1} =u$ and
  $u_2 = u_s =  \psi (xe^{\mu (p)t} ) $
 and obtain for the solution $u(x,t)$ of the problem
  \eqref{eq4.6}, \eqref{eq9.1}:
  \begin{equation}\label{eq9.5}
  \lim_{t\to\infty} u (xe^{-\mu t},t) =
  \begin{cases}
  1&\text{if }\ \mu >\mu (p)\\
  \psi (x)&\text{if }\ \mu = \mu (p)\\
  0&\text{if }\ \mu (p) > \mu > \mu (p +\delta)\ ,
  \end{cases}
  \end{equation}
  with sufficiently small $\delta >0$.

  We see that $\psi (x) = w(x^p)$, where $w(x)$ has all
  properties described in Theorem~\ref{theorem8.3}.
  The equalities \eqref{eq9.5} explain the exact meaning of
  the approximate identity,
  \begin{equation}\label{eq9.6}
  u(x,t) \approx \psi ( xe^{\mu (p)t})\ ,\qquad
  t\to \infty\ ,\ xe^{\mu (p)t} = \text{const.}\ ,
  \end{equation}
  that we call self-similar asymptotics.

In particular,  the following statement holds.
  \begin{proposition}\label{prop9.1}
  The solution $u(x,t)$ of the problem \eqref{eq4.6}, \eqref{eq9.1},
  with $\Gamma$ given in Eqs.~\eqref{eq4.3}, satisfies either one
of  the following
  limiting identities:
 \begin{itemize}
 \item[{\bf (1)}]
 \begin{equation}\label{eq9.3}
 \lim_{t\to \infty} u(xe^{-\mu t}, t) = 1\ ,\qquad x\ge 0\ ,
   \end{equation}
   for any $\mu >\mu (p_0)$.  if $p\ge p_0$ for the initial
data \eqref{eq9.1}\, ,
   \item[{\bf (2)}] ~Eqs.~\eqref{eq9.5} provided $0<p<p_0$.
 \end{itemize}
The  convergence in Eqs.~\eqref{eq9.3}, \eqref{eq9.5} is
uniform on
  any bounded interval $0\le x\le R$, and
 \begin{equation*}
u ( xe^{\mu (p)t},t) -\psi(x) = O(x^{p+\ep}) e^{-\beta(p,\ep)t}\, ,\qquad
\beta(p,\ep)= (p+\ep)\,(\mu(p) -\mu(p+\ep) \,),
\end{equation*}
for   $0<p<p_0$ and $0<\ep<p_0-p$. \end{proposition}

\begin{remark} There is a connection
between  self-similar asymptotics and non-linear wave propagation.
It is easy to see that  self-similar asymptotics becomes more
  transparent in logaritheoremic variables
  \begin{equation*}
  y = \ln x\ ,\quad u(x,t) = \hat u(y,t)\ ,\quad \psi (x,t) = \hat\psi(y,t)
  \end{equation*}
  Thus, Eq.~\eqref{eq9.6} becomes
  \begin{equation}\label{eq9.7}
  \hat u(y,t) \approx \hat\psi (y + \mu (p)t)\ ,\qquad
  t\to \infty\ ,\quad y+\mu (p)t = \text{const. },
  \end{equation}
  hence, the self-similar solutions are simply nonlinear waves
  (note that $\psi (-\infty) =1$, $\psi (+\infty) =0$)
  propagating with constant velocities $c_p = -\mu (p)$ to
  the right if $c_p >0$ or to the left if $c_p <0$.
  If $c_p >0$ then the value $u(-\infty,t)=1$ is transported
  to any given point $y\in\real$ when $t\to\infty$.
  If $c_p <0$ then the profile of the wave looks more
  naturally for the  functions $\tilde u =1-\hat u$, $\tilde\psi=1-\psi$.

We conclude that Eq.~\eqref{eq4.6} can be considered in some sense as
  the equation for nonlinear waves.
  The self-similar asymptotics \eqref{eq9.7}
  means a formation of the traveling wave with a universal
  profile for a broad class of initial conditions.
 This is a purely non-linear phenomenon, it is easy to see
  that such asymptotics cannot occur in the particular case
  ($N=1$ in Eqs.~\eqref{eq4.3}) of the linear operator $\Gamma$.

\end{remark}

\vskip.5in
 \subsection{Distribution functions, moments and power-like tails}\label{sec:6.2}

 We have described above the general picture of behavior of the solutions
  $u(x,t)$ to the problem \eqref{eq4.6}, \eqref{eq9.1}.
  On the other hand, Eq.~\eqref{eq4.6} (in particular, its special
  case \eqref{eq3.2}) was obtained as the Fourier transform of
  the kinetic equation.
  Therefore we need to study in more detail the corresponding
  distribution functions.

Set  $u_0(x)$ in the problem
 \eqref{eq4.6} to be  an isotropic characteristic function of a
 probability measure in $\real^d$, i.e.,
 \begin{equation}\label{eq10.1}
 u_0(x) = \F [f_0] = \int_{\real^d} dv\, f_0 (|v|) e^{-ik\cdot v}\ ,
 \qquad k\in \real^d\ ,\ x= |k|^2\ ,
 \end{equation}
 where $f_0$ is a generalized positive function normalized
 such that $u_0(0)=1$ (distribution function).
 Let $U$ be a closed unit ball  in the $B=C(\real_+)$ as defined
 in \eqref{eq4.10}.

 Then,  we can apply all results of  Section~5,
 and conclude that there
 exists a distribution function $f(v,t)$, $v\in\real^d$, satisfying
 Eq.~\eqref{eq2.1}, such that
 \begin{equation}\label{eq10.2}
 u(x,t) = \F [f(\cdot,t)]\ ,\qquad x = |k|^2\ ,
 \end{equation}
 for any $t\ge 0$, and a  similar conclusion can be obtain if
 we assume the Laplace
 (instead of Fourier) transform in Eqs.~\eqref{eq10.1}.

 Then there exists a distribution function $f(v,t)$, $v>0$,
 such that
 \begin{equation}\label{eq10.3}
 u(x,t) = \L [f(\cdot,t)] = \int_0^\infty dv\, f(v,t) e^{-xv}\ ,\qquad
 u(0,t) = 1\ ,\ \ x\geq 0\,,\ \ t\ge 0\ ,
 \end{equation}
 where $u(x,t)$ is the solution of the problem \eqref{eq4.6}
 constructed in Theorem~\ref{theorem5.2} and Lemma~\ref{lemma5.3}.

 The approximate equation \eqref{eq9.6} in terms of
 distribution functions \eqref{eq10.2} reads
 \begin{equation}\label{eq10.4}
 f(|v|,t) \simeq e^{-\frac{d}2 \mu(p)\, t} F_p (|v| e^{-\frac12 \mu(p)\,t})\ ,
 \quad t\to \infty\ ,\ |v| e^{-\frac12 \mu(p)\, t} = \text{const.}\ ,
 \end{equation}
 where $F_p (|v|)$ is a distribution function such that
 \begin{equation}\label{eq10.5}
 \psi_p (x) = \F [F_p]\ ,\qquad x= |k|^2\ ,
 \end{equation}
 with $\psi_p $ given by
 \begin{equation}\label{eq9.4}
  u_s (x,t) = \psi (xe^{\mu (p)t} )
   \end{equation} (the notation $\psi_p$
 is used in order to stress that $\psi$ defined in \eqref{eq9.4},
  depends on $p$).
 The factor $1/2$ in Eqs.~\eqref{eq10.4} is due to the notation
 $x= |k|^2$.
 Similarly, for the Laplace transform, we obtain
 \begin{equation}\label{eq10.6}
 f(v,t) \simeq e^{-\mu (p)t} \Phi_p (ve^{-\mu (p)t})\ ,\qquad
 t\to\infty\ ,\ ve^{-\mu (p)t} = \text{const.}\ ,
 \end{equation}
 where
 \begin{equation}\label{eq10.7}
 \psi_p(x) = \L [\Phi_p]\ .
 \end{equation}

 The positivity and some other properties of $F_p (|v|)$ follow from
 the fact that $\psi_p (x) = w_p(x^p)$, where $w_p(x)$ satisfies
 Theorem~\ref{theorem8.3}.
 Hence
 \begin{equation}\label{eq10.8}
 \psi_p (x) = \int_0^\infty d\tau\, R_p (\tau) e^{-\tau x^p}\ ,
 \qquad \int_0^\infty d\tau\, R_p(\tau)
 = \int_0^\infty d\tau\, R_p (\tau) \tau = 1\ ,
 \end{equation}
 where $R_p (\tau)$, $\tau \ge 0$, is a non-negative generalized
 function (of course, both $\psi_p$ and $R_p$ depend on $p$).


 In particular we can conclude that
 the self-similar asymptotics \eqref{eq10.4} for any
 initial data $f_0\ge 0$ occurs if $p_0 >1 $,
 otherwise it occurs for $p\in (0,p_0) \subset (0,1)$.
 Therefore, for any spectral function $\mu (p)$ (Fig.1), the
 approximate equality \eqref{eq10.4} holds for sufficiently
 small $0<p\le 1$.
In addition,
 \begin{equation*}
 m_2 = \int_{\real^d} dv\, f_0 (|v|) |v|^2 <\infty \ \text{ if }\ p=1
 \end{equation*}
 and $m_2=\infty$  if $p<1$.
 Similar conclusions can be made for the Laplace transforms.

 The positivity of $F(|v|)$ in Eqs.~\eqref{eq10.5}--\eqref{eq10.7}
 follows from the integral representation \eqref{eq10.8} with $p\le1$, since
 it is well-known that
 \begin{equation*}
 \F^{-1} (e^{-|k|^{2p}} ) >0\ ,\qquad \L^{-1} (e^{-x^{2p}}) >0
 \end{equation*}
 for any $0<p\le 1$ (the so-called infinitely divisible distributions
 \cite{Fe}).
 Thus, Eqs.~\eqref{eq10.8} explains the connection of the  self-similar
 solutions of generalized Maxwell models with infinitely
 divisible distributions.

 Using standard formulas for the inverse Fourier (Laplace)
 transforms, denote by ($d= 1,2,\ldots$ is fixed)
 \begin{equation}\label{eq10.10}
 \begin{split}
 M_p (|v|)
 & = \frac1{(2\pi)^d} \int_{\real^d} dk e^{-|k|^{2p} +ik\cdot v}\ ,\\
 \noalign{\vskip6pt}
 N_p (v) & = \frac1{2\pi i}\int_{a-i\infty}^{a+i\infty} dx\,
 e^{-x^p +xv}\ ,\qquad 0< p\le 1\ ,
 \end{split}
 \end{equation}
 we obtain  the self-similar solutions (distribution functions) are
 given in Eqs.~\eqref{eq10.6}, \eqref{eq10.8} (right hand sides),
 by
 \begin{equation}\label{eq10.11}
 \begin{split}
 F_p (|v|) & = \int_0^\infty d\tau\, R_p (\tau) \tau^{-\frac{d}{2p}} M_p
 (|v| \tau^{-\frac1{2p}})\ ,\\
 \noalign{\vskip6pt}
 \Phi_p (v) & = \int_0^\infty d\tau\, R_p (\tau) \tau^{-\frac1p} N_p
 (v\tau^{-\frac1p})\ ,\qquad v \ge 0\ ,\ 0<p\le 1\ .
 \end{split}
 \end{equation}
 Note that $M_1(|v|)$ is the standard Maxwellian in $\real^d$.
 The functions $N_p (v)$ \eqref{eq10.10} are studied in detail
 in the  literature \cite{Fe,Lu}.
 Thus, for given $0<p\le 1$, the kernel $R(\tau)$, $\tau\ge0$,
 is the only unknown function that is needed to describe the
  distribution functions $F(|v|)$ and $\Phi (v)$.
 See \cite{BoCeGa} for a  study of $R(\tau)$ in more detail.

Now, from  \eqref{eq5.3}, \eqref{eq6.6},
and recalling  $\mu = \mu (1)$,
we can show the moments  equation can be written in the form
\begin{equation}\label{eq10.15}
(s\mu (1) - \lambda (s) + 1) m_s = \sum_{n=2}^N \alpha_n I_n (s)\ ,
\end{equation}
where
\begin{equation}\label{eq10.16}
\begin{split}
&I_n (s) = \int_{\real_+^n}\mkern-18mu  da_1\ldots da_n A (a_1,\ldots,a_n)
\int_{\real_+^n}  \mkern-18mu  d\tau_1\ldots, d\tau_n\; g_n^{(s)}
(a_1\tau_1,\ldots, a_n \tau_n) \prod_{j=1}^n R(\tau_j)\\
\noalign{\vskip6pt}
&g_n^{(s)} (y_1,\ldots, y_n)
= \bigg( \sum_{k=1}^n y_k\bigg)^s
- \sum_{k=1}^n y_k^s\ ,\qquad n =1,2,\ldots\ .
\end{split}
\end{equation}
and, due to the properties of $R(\tau)$, one gets
 $g_1^{(s)} =0$ for any $s\ge0$ and  $m_0 = m_1=1$.

Our aim is to study the moments $m_s$ defined in Eq.\eqref{eq10.3},
 for $s>1$, on
the basis of Eq.~\eqref{eq10.15}. The approach is similar to the one
used in \cite{To-05} for a simplified version of Eq.~\eqref{eq10.15}
with $N=2$. The main results are formulated below in terms of the
spectral function $\mu (p)$ (see Fig.1) under assumption that
$p_0>1$.

\begin{proposition}\label{prop10.1}
$\quad$
\begin{itemize}
\item[{\bf [i]}] If the equation $\mu (s) = \mu (1)$ has the only solution $s=1$,
then $m_s <\infty$ for any $s>0$.
\item[{\bf [ii]}] If this equation has two solutions $s=1$ and $s= s_* >1$,
then $m_s <\infty$ for $s<s_*$ and $m_s =\infty$ for $s>s_*$.
\item[{\bf [iii]}] $m_{s_*} <\infty$ only if $I_n (s_*)=0$ in Eq.~\eqref{eq10.15}
for all $n=2,\ldots,N$.
\end{itemize}
\end{proposition}

Now we can draw some conclusions concerning the moments of
the distribution functions \eqref{eq10.11} as follows.
Denote by
\begin{equation*}
\begin{split}
m_s (\Phi_p) & = \int_0^\infty dv\, \Phi_p (v) v^s\ ,\qquad
m_s (R_p) = \int_0^\infty d\tau\, R_p (\tau) \tau^s\ ,\\
\noalign{\vskip6pt}
m_{2s} (F_p) & = \int_{\real^d} dv\, F_p (|v|) |v|^{2s}\ ,\qquad
s>0\ ,\ 0< p\le 1\ ,
\end{split}
\end{equation*}
and use similar notations for $N_p (v)$ and $M_p(|v|)$ in
Eqs.~\eqref{eq10.11}.
Then, by formal integration of Eqs.~\eqref{eq10.11}, we obtain
\begin{equation*}
\begin{split}
m_s (\Phi_p) & = m_s (N_p) m_{s/p} (R_p)\\
\noalign{\vskip6pt}
m_{2s} (F_p) & = m_{2s} (M_p) m_{s/p} (R_p)\ ,
\end{split}
\end{equation*}
where $M_p$ and $N_p$ are given in Eqs.~\eqref{eq10.10} and we show that
finite only for $s<p<1$.

In the remaining case $p=1$  {\em all\/}
moments of functions
\begin{gather*}
M_1 (|v|) = (4\pi)^{-d/2} \exp \Big[ - \frac{|v|^2}4\Big] \ ,\qquad
v\in \real^d\ ;\\
\noalign{\vskip6pt}
N_1 (v) = \delta (v-1)\ ,\qquad v\in \real_+\ ,
\end{gather*}
are finite.
Therefore,
 everything depends on moments of $R_1$ in with $p=1$, so it only needs
to apply Proposition~\ref{prop10.1}.

In particular, the following statement  holds for the moments of
the distribution functions \eqref{eq10.4}, \eqref{eq10.6}.

\begin{proposition}\label{prop10.3}
$\quad$
\begin{itemize}
\item[{\bf [i]}] If $0<p<1$, then $m_{2s} (F_p)$ and $m_s (\Phi_p)$
are finite if and only if $0<s<p$.
\item[{\bf [ii]}] If $p=1$, then Proposition~\ref{prop10.1} holds for
$m_s = m_{2s} (F_1)$ and for $m_s = m_s (\Phi_1)$.
\end{itemize}
\end{proposition}

\begin{remark}
Proposition~\ref{prop10.3} can be interpreted in other words:
the distribution functions $F_p(|v|)$ and
$\Phi_p(v)$, $0<p\le 1$, can have finite moments of all orders
in the only case when two conditions are satisfied
\begin{itemize}
\item[{\bf{\rm (1)}}] $p=1$,  and
\item[{\bf{\rm (2)}}] the equation $\mu (s)  = \mu (1)$ (see Fig.1)
has the unique solution $s=1$.
\end{itemize}
In all other cases, the maximal order $s$ of finite moments
$m_{2s} (F_p)$ and $m_s (\Phi_p)$ is bounded.
\end{remark}

This fact  means that the distribution functions
$F_p$ and $\Phi_p$ have power-like tails.


 \subsection{Applications to the conservative or dissipative Boltzmann equation}\label{sec:6.3}

We recall the three specific Maxwell models {\bf (A)}, {\bf (B)}, {\bf (C)}
 of the  Boltzmann equation from section 2. Our goal in this section
is to study isotropic solutions  $f( |v |, t), v\in\real^d$,
 of Eqs.~\eqref{eq2.2}, \eqref{eq2.4}, and
 \eqref{eq2.5} respectively.
 All three cases are considered below from a  unified point of view.
 First we perform the Fourier transform and denote
\begin{equation}\label{eq11.1}
\begin{split} u(x,t) = \F [f(|v|,t)] = \int_{\real^{d}} dv\, f(|v|,t)
e^{-ik\cdot v}, \quad x=|k|^2, \ \ u(0,t)=1.
\end{split}
\end{equation}
It was already said at the beginning of section~4 that $u(x,t)$
satisfies (in all three cases) Eq.~\eqref{eq4.2}, where $N=2$
 and all notations are given in Eqs.\eqref{eq4.5},
 \eqref{eq3.2}, \eqref{eq3.6}-\eqref{eq3.9}.
 Hence, all results of our general theory are applicable to these specific
 models. In all three cases  {\bf (A)}, {\bf (B)}, {\bf (C)}
 we assume that the initial distribution function
\begin{equation}\label{eq11.2}
 f(|v|,0)= f_0(|v|) \geq 0\, ,\qquad  \int_{\real^{d}} dv\, f_0(|v|)=1\, ,
\end{equation}
and the corresponding characteristic function
\begin{equation}\label{eq11.3}
u(0,t)= u_0(x) = \F[f_0(|v|)]\, ,\quad  x=|k|^2 \, ,
\end{equation}
are given. Moreover, let $u_0(x)$ be such that
\begin{equation}\label{eq11.4}
 u_0(x) = 1-\alpha x^p + O(x^{p+\ep}),\qquad x\to 0, \qquad  0<p\leq 1,
\end{equation}
with some $\alpha>0$ and $\ep>0$.
We distinguish below the initial data with finite energy (second moment)
\begin{equation}\label{eq11.5}
 E_0 = \int_{\real^{d}} dv\,|v|^2 f_0(|v|) <\infty
\end{equation}
implies $ p=1$ in Eqs.~\eqref{eq11.4}
 and the in-data with infinite energy $E_0=\infty$.
If  $p<1$
 in Eqs.~\eqref{eq11.4} then
\begin{equation}\label{eq11.6}
m_q^{(0)}= \int_{\real^{d}} dv\, f_0(|v|) |v|^{2q} <\infty
\end{equation}
only for $q\leq p<1$ (see \cite{Fe, Lu}).

Also, the case $p>1$ in Eqs.~\eqref{eq11.4}  is not possible for
$f_0(|v|)\geq 0$.

In addition, note that the coefficient $\alpha>0$ in Eqs.~\eqref{eq11.4}
can always be changed to $\alpha=1$ by the scaling transformation
$\tilde x =\alpha^{1/p} x$. Then, without loss of generality,
 we set $\alpha=1$ in \eqref{eq11.4}.

Since it is known that the operator $\Gamma(u)$ in all three cases
belongs to the class \eqref{eq4.3}, we can apply
Theorem~\ref{theorem8.3}
 and state that self-similar solutions of Eq.~\eqref{eq4.2}
are given by
\begin{equation}\label{eq11.7}
u_s(x,t) = \Psi(x\, e^{\mu(p) t}), \ \ \ \Psi(x)=w(x^p)\, ,
\end{equation}
where $w(x)$ is given in theorem~\ref{theorem8.3} and $0<p<p_0$ (the
spectral function $\mu(p)$, defined in \eqref{spec-func}, and its
critical point $p_0$ depends on the specific model.)

According to Sections 9-10, we just need to find the spectral function
 $\mu(p)$. In order to do this we first define the linearized
 operator $L=\Gamma'(1)$ for $\Gamma(u)$ given in
Eqs.~\eqref{eq4.3}, \eqref{eq4.5}.
One should be careful at this point since
$A_2(a_1,a_2)$ in Eqs.~\eqref{eq4.5}  is not symmetric and therefore
Eqs.~\eqref{eq4.13}
cannot be used. A straight-forward computation leads to
 \begin{equation}\label{eq11.8}
 \begin{split}
 L\, u(x) & = \int_0^1 ds\, G(s) (\, u(a(s)x) + u(b(s)x)\, )
 +
 \int_0^1 ds H(s)  u(c(s)x) \ ,
 \end{split}
 \end{equation}
in the notation \eqref{eq3.6} -  \eqref{eq3.9}. Then,
 the eigenvalue $\lambda(p)$ is given by
 \begin{equation}\label{eq11.9}
 \begin{split}
L\, x^p &  =\lambda(p)\, x^p\ \ \ \text{which implies}\\
 \lambda(p) & = \int_0^1 ds\, G(s) \left\{  (a(s))^p + (b(s))^p
 \right\}+
 \int_0^1 ds H(s)  (c(s))^p \ ,
 \end{split}
 \end{equation}
and the spectral function  \eqref{spec-func} reads
 \begin{equation}\label{eq11.10}
 \begin{split}
\mu(p)=\frac{\lambda(p)-1 }{p}\, .
 \end{split}
 \end{equation}
Note that the normalization \eqref{eq4.1} is assumed.

At that point we consider the three models {\bf (A), (B), (C) }
separately and apply Eqs.~\eqref{eq11.9} and
 \eqref{eq11.10} to each case.

\noindent {\bf (A) } Elastic Boltzmann Equation~\eqref{eq2.2} in $\real^d, d\geq 2$.
By using Eqs.~\eqref{eq3.6}, \eqref{eq3.7}, and \eqref{eq4.1}
we obtain
 \begin{equation}\label{eq11.11}
 \begin{split}
 \lambda(p) = \int_0^1 ds\, G(s) ( s^p + (1-s)^p ) \, ,
\qquad G(s) =A_d\, g(1-2s) [s(1-s)]^{\frac{d-3}2}\, ,
 \end{split}
 \end{equation}
where the normalization constant $A_d$ is such that Eq.~\eqref{eq4.1}
is satisfied with $H=0$. Then
 \begin{equation}\label{eq11.12}
 \begin{split}
 \mu(p) = \frac1p\int_0^1 ds\, G(s) ( s^p + (1-s)^p - 1) \, , p>0.
 \end{split}
 \end{equation}
It is easy to verify that $p\,\mu(p) \to 1$ as $p\to 0$, $\mu(p) \to
0$ as $p\to\infty$, and
 \begin{equation}\label{eq11.13}
 \begin{split}
 &\mu(p) >0 \ \ \text{if}\ p<1\,; \qquad  \qquad\mu(p) <0
\ \ \text{if}\ p>1\,;\\
&\mu(1)=0\, ,\ \ \ \ \ \ \ \qquad \qquad \mu(2)=\mu(3)= -\int_0^1
ds\, G(s) \, s (1-s) \, .
\end{split}
 \end{equation}
 Hence, $\mu(p)$ in this case is similar to the function shown on Fig.1
{\bf b)} with $2<p_0<3$ and such that $\mu(1)=0$. Then the
self-similar
 asymptotics hold all $0<p<1$.

\noindent {\bf (B) }{\bf Elastic Boltzmann Equation in the presence of a
thermostat~\eqref{eq2.4} in $\real^d, d\geq2$.}
We consider just the case of a cold thermostat with $T=0$ in
Eq.~\eqref{eq3.2}, since the general case $T>0$ can be considered after
 that with the help of \eqref{eq2.11}.
Again, by using Eqs.~\eqref{eq3.6}, \eqref{eq3.7}, and \eqref{eq4.1}
we obtain
 \begin{equation}\label{eq11.14}
 \begin{split}
 \lambda(p) &= \int_0^1 ds\, G(s) ( s^p + (1-s)^p ) +
\theta\, \int_0^1 ds\, G(s) ( 1- \frac{4m}{(1+m)^2} )^p ,\\
 G(s) & = \frac{1}{1+\theta}\, A_d\, g(1-2s) [s(1-s)]^{\frac{d-3}2}\, ,
 \end{split}
 \end{equation}
with the same  constant $A_d$ as in Eq.~\eqref{eq11.11}. Then
 \begin{equation}\label{eq11.15}
 \begin{split}
 \mu(p) &= \frac{1}{p}\int_0^1 ds\, G(s)
( \, s^p + (1-s)^p - \theta(1-\beta s)^p
-(1+\theta) \, )\, , \\
 \beta &= \frac{4m}{(1+m)^2}\, , \ \  p>0,
 \end{split}
 \end{equation}
and therefore, as in the previous case {\bf A)}, $p\,\mu(p) \to 1$
as $p\to 0$, $\mu(p) \to 0$ as $p\to\infty$, with
 \begin{equation}\label{eq11.16}
 \begin{split}
\  &\mu(1)=  -\theta\, \beta\,\int_0^1 ds\, G(s) \, s  \, .
\end{split}
 \end{equation}
which again verifies that  $\mu(p)$
is of the same kind as in the elastic case {\bf A)} and shown on
Fig.1~{(b)}.
A position of the critical point $p_0$ such that $\mu'(p_0)=0$
(see Fig.1~{(b)})  depends on $\theta$. It is important to distinguish
two cases: {\bf 1)} $p_0>1$ and  {\bf 2)} $p_0<1$. In case   {\bf 1)} any
 non-negative initial data \eqref{eq11.2} has the self-similar asymptotics.
In  case  {\bf 2)} such asymptotics holds just for in-data with infinity
 energy satisfying Eqs.~\eqref{eq11.4} with some $p<p_0<1$. A simple
 criterion to separate  the two cases follows directly from   Fig.1~{(b)}:
it is enough to check the sign of $\mu'(1)$. If
 \begin{equation}\label{eq11.17}
\mu'(1) =\lambda'(1) -\lambda(1) +1 <0
 \end{equation}
in the notation of Eqs. \eqref{eq11.14}, then $p_0>1$
and the self-similar asymptotics hold for any non-negative initial data.

The inequality \eqref{eq11.17} is equivalent to the following condition on
 the positive coupling constant $\theta$
 \begin{equation}\label{eq11.18}
 \begin{split}
0 < \theta<\theta_* = - \frac{\int_0^1 ds\, G(s)\ (\, s \log s +
(1-s)\log(1-s)\, )}
{ \int_0^1 ds\, G(s)\ (\, \beta \, s  + (1-\beta s)\log(1-\beta s)\, ) }\, .
\end{split}
 \end{equation}
The right hand side of this inequality is positive and independent on the
 normalization of $G(s)$, therefore it does not depend on $\theta$
(see Eq.~\eqref{eq11.15}. We note that a new class of exact self-similar
solutions to Eq.~\eqref{eq2.4} with finite energy was recently found in
\cite{BG-06} for $\beta=1, \theta=4/3$ and $G(s)=const.$
A simple calculation of the
integrals in \eqref{eq11.18} shows that $\theta_*=2$ in
that case, therefore the criterion \eqref{eq11.17} is fulfilled
for the exact solutions from \cite{BG-06} and they are asymptotic for a wide
 class of initial data with finite energy.
Similar conclusions can be made in the same way about exact positive
self-similar solutions with infinite energy
 constructed in \cite{BG-06}. Note that the
 inequality \eqref{eq11.17}  shows the non-linear character of the
self-similar asymptotics: it holds  unless the linear
 term in  Eq.~\eqref{eq2.4}
 is `too large'.

\noindent {\bf (C) } Inelastic Boltzmann Equation~\eqref{eq2.5} in
$\real^d$. Equations \eqref{eq3.9} and \eqref{eq4.1} lead to
 \begin{equation*}
\label{eq11.19n}
 \begin{split}
 &\qquad\lambda(p) = \int_0^1 ds\, G(s) ( (a\,s)^p + (1-b\,s)^p ) \, ,\\
&\text{where}\\
&\qquad G(s) = C_d\,  (1-s)^{\frac{d-3}2}\, ,\qquad a=\frac{(1+e)^2}4
   \, ,\
\ \ \ b=\frac{(1+e)(3-e)}4\, ,
 \end{split}
 \end{equation*}
with such constant $C_d$ that Eq.~\eqref{eq4.1} with $H=0$ is fulfilled.
Hence
 \begin{equation}\label{eq11.19}
 \begin{split}
 \mu(p) = \frac1p\int_0^1 ds\, G(s) (\,  (a\,s)^p + (1-b\,s)^p  - 1) \, ,
\qquad p>0\, ,
 \end{split}
 \end{equation}
and once more as in the previous two cases,  $p\,\mu(p) \to 1$ as
$p\to 0$, $\mu(p) \to 0$ as $p\to\infty$, with now
 \begin{equation*}
 \begin{split}
&\mu(1)=-\frac{1-e^2}4\int_0^1 ds\, G(s) \,s \, .
\end{split}
 \end{equation*}
Thus, the same considerations lead to the shape of $\mu(p)$ shown in
 Fig.1~{(b)}.
The inequality  \eqref{eq11.17} with $\lambda(p)$ given in Eqs.
\eqref{eq11.19} was proved in \cite{BCT-03} (see Eqs. (4.26) of \cite{BCT-03},
where the notation is slightly different from ours). Hence, the
inelastic Boltzmann equation~\eqref{eq2.5}  has self-similar asymptotics for
any restitution coefficient $0<e<1$ and any non-negative initial data.

\medskip

Hence, the spectral function $\mu(p)$ in all three cases above is such
 that $p_0>1$ provided the inequality \eqref{eq11.18} holds for the model
{\bf (B)}.

Therefore, according to our general theory, all `physical' initial
conditions \eqref{eq11.2} satisfying Eqs. \eqref{eq11.4} with any
$0<p\leq 1$ lead to self-similar asymptotics. Hence, the main
properties of the solutions $f(v,t)$ are qualitatively similar for
all three models {\bf (A), (B)} and {\bf (C)}, and  can be described
in one unified statement: Theorem~\ref{theorem11.1} below.

Before we formulate such general statement, it is worth to clarify one
 point related to a special value $0<p_1\leq 1$ such that $\mu(p_1)=0$.
The reader can see that on Fig.1~{(b)} that the unique root of this
equation exits  for all models {\bf (A), (B),}{\bf (C)}
 since $\mu(1)=0$ in the case  {\bf (A)} (energy conservation) , and
$\mu(1)<0$ in  cases  {\bf (B)} and  {\bf (C)} (energy dissipation).
If $p=p_1$ in Eqs. \eqref{eq11.4} then the self-similar solution
\eqref{eq11.7} is simply a stationary solution of Eq. \eqref{eq4.2}.
Thus, the time relaxation to the non-trivial ($u\neq 0,1$)
stationary solution is automatically included in
Theorem~\ref{theorem11.1} as a particular  case of self-similar
asymptotics.

Thus we consider simultaneously Eqs.~\eqref{eq2.2}, \eqref{eq2.4},
\eqref{eq2.5}, with the initial condition \eqref{eq11.2} such that
Eq. \eqref{eq11.4} is satisfied with some $0<p\leq 1$, $\ep >0$
and $\alpha=1$. We also assume that $T=0$ in Eq. \eqref{eq2.4} and the
 coupling parameter $\theta>0$ satisfies the condition \eqref{eq11.18}.

In the following Theorem~\ref{theorem11.1},  the solution $f(|v|,t)$
is understood in each case as a generalized density of probability
measure in $\real^d$ and the convergence  $f_n\to f$ in the sense of
weak
 convergence of probability measures.

\begin{theorem}\label{theorem11.1} The following two statements hold

\begin{itemize}
\item[{\bf [i]}]  There exists a unique (in the class of probability measures)
solution $f(|v|,t)$ to each of Eqs.~\eqref{eq2.2}, \eqref{eq2.4},
\eqref{eq2.5} satisfying the initial condition  \eqref{eq11.2}. The solution
 $f(|v|,t)$  has self-similar asymptotics in the following sense:

For any given $0<p\leq 1$ in  Eqs.~\eqref{eq11.4}
there exits a unique non-negative self-similar solution
 \begin{equation}\label{eq11.20}
 \begin{split}
 f_s^{(p)}(|v|,t) = e^{-\frac{d}2\mu(p)\,t} F_p(|v|e^{-\frac12{\mu(p)}\, t})\ ,
 \end{split}
 \end{equation}
such that
 \begin{equation}\label{eq11.21}
 \begin{split}
 e^{\frac{d}2\mu(p)\, t} f(|v| e^{-\frac12 \mu(p)\,t},t) \
\to_{t\to\infty}  F_p(|v|)\ ,
 \end{split}
 \end{equation}
where $\mu(p)$ is given in Eqs.~\eqref{eq11.12}, \eqref{eq11.15},
\eqref{eq11.19}, respectively, for each of the three models.
\item[{\bf [ii]}] Except for the special case of the Maxwellian
 \begin{equation}\label{eq11.22}
 \begin{split}
F_1(|v|)=
M(|v|) =(4\pi)^{-d/2} e^{-\frac{|v|^2}4}
 \end{split}
 \end{equation}
for Eq.~\eqref{eq2.2} with $p=1$ in Eq.~\eqref{eq11.4} (note that $\mu(1)=0$ in this case), the function $ F_p(|v|)$ does not have finite moments of all orders.
If $0<p<1$, then
\begin{equation}\label{eq11.23}
m_q = \int_{\real^d} dv\, F_p(|v|) |v|^{2q} < \infty \qquad\text{only for  }
0<q<p\, .
 \end{equation}
If $p=1$ in the case of Eqs.~\eqref{eq2.4},
\eqref{eq2.5}, then $m_q <\infty$ only for  $0<q<p_*$, where $p_*>1$ is the unique maximal root of the equation $\mu(p_*)=\mu(1)$, with   $\mu(p)$ given in
Eqs.~\eqref{eq11.4}, \eqref{eq11.19} respectively.
\end{itemize}
\end{theorem}

In addition,  we also we obtain the following corollary.

\begin{corollary}\label{cor11.2}
Under the same  conditions of Theorem~\ref{theorem11.1}, the
following two statements hold.
\begin{itemize}
\item[{\bf [i]}] The rate of convergence in Eq. \eqref{eq11.21} is
 characterized in terms of the corresponding characteristic functions
 in Proposition~\ref{prop9.1}.
\item[{\bf [ii]}] The function $ F_p(|v|)$
admits the integral representation \eqref{eq10.11} through infinitely
divisible distributions  \eqref{eq10.10}.
\end{itemize}
\end{corollary}
\begin{proof}
It is enough to note that all results of section 9 and 10 are valid,
 in particular, for Eqs.~\eqref{eq2.2}, \eqref{eq2.4},
\eqref{eq2.5}.
\end{proof}

Finally, we mention that  the statement similar to
theorem~\ref{theorem11.1},
 can be easily derived  from general results of sections 7.1 and 7.2 in
the case
 of $1-d$ Maxwell models introduced in \cite{To-05}, \cite{BN05}
for applications to economy models (Pareto tails, etc.). The only difference
 is that the `kinetic' equation can be transformed to its canonical form
 \eqref{eq4.2}-\eqref{eq4.3} by the Laplace transform as discussed in
 section 10, and that the spectral function $\mu(p)$ can have in this case
 any of the four kind of behaviors shown in Fig.1. The only remaining
 problem for any such $1-$dimensional models is to study them for their  specific
function $\mu(p)$, and then to apply
 Propositions~\ref{prop9.1}, \ref{prop10.1}
 and \ref{prop10.3}.

Thus, the general theory developed in this chapter is applicable to
all
 existing multi-dimensional isotropic Maxwell models and to $1\,d$ models as well.

\begin{acknowledgement}
The first author was supported by  grant 621-2003-5357 from the
Swedish Research Council (NFR). The research of the  second author
was supported by MIUR of Italy.  The third author
 has been partially supported by NSF under grant
DMS-0507038. Support from the Institute for Computational
Engineering and Sciences at the University of Texas at Austin is
also gratefully acknowledged. \end{acknowledgement}


 \end{document}